\PassOptionsToPackage{table,prologue,dvipsnames}{xcolor}
\documentclass[manuscript, nonacm]{acmart}
\usepackage{xcolor}
\usepackage[utf8]{inputenc}
\usepackage{etoolbox}
\usepackage{amsmath}
\usepackage{mathtools}
\usepackage{tabularx}
\usepackage{xltabular}
\usepackage{wrapfig,booktabs}
\usepackage{enumitem}
\usepackage{makecell}
\usepackage{multirow}
\usepackage{soul,color}
\usepackage{xspace}
\usepackage{graphicx}
\usepackage{subcaption}
\usepackage{cleveref}
\usepackage{float}
\usepackage{placeins}
\usepackage{dashrule}
\usepackage[final, markup=underlined, commandnameprefix=ifneeded, defaultcolor=BrickRed]{changes}
\definecolor{gray}{gray}{0.9}

\newcommand{\mycomment}[1]{}
\newcommand{\jb}{JetBrains\xspace} % JB mention
\newcommand{\ninters}{38\xspace} % Number of interviews
\newcommand{\nprobs}{42\xspace} % Number of problems
\newcommand{\surveyOne}{``Undesirable Patterns Overview''\xspace} % Name of the first survey
\newcommand{\nsurveyOne}{436\xspace} % Number of first survey participants
\newcommand{\surveyTwo}{``Evaluation of Tool Suggestions''\xspace} % Name of the second survey
\newcommand{\nsurveyTwo}{968\xspace} % Number of second survey participants
\newcommand{\cdps}{collective development problems\xspace}
\newcommand{\cdp}{collective development problem\xspace}
\newcommand{\ups}{undesirable patterns\xspace}
\newcommand{\up}{undesirable pattern\xspace}
\newcommand{\observation}[1]{\refstepcounter{observation}
	\begin{center}
		\framebox{
			\begin{minipage}{0.93\columnwidth}
				{} #1
			\end{minipage}
		}
	\end{center}
}

\title{What Could Possibly Go Wrong: Undesirable Patterns in Collective Development}
\author{Mikhail Evtikhiev}
\authornote{Both authors contributed equally to this research.}
\email{mikhail.evtikhiev@jetbrains.com}
\orcid{0000-0002-1004-8943}
\affiliation{%
  \institution{JetBrains Research}
  \city{Limassol}
  \country{Cyprus}
}
\author{Ekaterina Koshchenko}
\authornotemark[1]
\email{ekaterina.koshchenko@jetbrains.com}
\orcid{0000-0003-3375-037X}
\affiliation{%
  \institution{JetBrains Research}
  \city{Amsterdam}
  \country{The Netherlands}
}
\author{Vladimir Kovalenko}
\email{vladimir.kovalenko@jetbrains.com}
\orcid{0000-0001-5880-7323}
\affiliation{%
  \institution{JetBrains Research}
  \city{Amsterdam}
  \country{The Netherlands}
}
\date{December 2023}
\begin{document}
\definechangesauthor[color=BrickRed]{rev}
\begin{abstract}
Software development, often perceived as a technical endeavour, is fundamentally a social activity requiring collaboration among team members.
Acknowledging this, the software development community has devised strategies to address possible collaboration-related shortcomings.
Various studies have attempted to capture the social dynamics within software engineering. 
In these studies, the authors developed \added{methods} to identify numerous teamwork issues and proposed various approaches to address them.
However, certain teamwork issues remain unstudied, necessitating a comprehensive bottom-up exploration \added{from practitioner's perceptions to common patterns}. 
This paper introduces the concept of \ups in collective development, referring to potential teamwork problems that may escalate if unaddressed.
Through \added{38} in-depth exploratory interviews, we identify and classify \nprobs patterns, revealing their origins and consequences. 
Subsequent surveys, \added{436 and 968 participants each}, explore the significance and frequency of the \ups, and evaluate potential tools and features to manage these patterns. 
The study contributes a nuanced understanding of \ups, evaluating their impact and proposing pragmatic tools and features for industrial application. 
The findings provide a valuable foundation for further in-depth studies and the development of tools to enhance collaborative software engineering practices.
\end{abstract}
\maketitle

\newcounter{observation}
\section{Introduction}\label{sec:intro}
While software development was historically considered to be a technical activity~\cite{espinosa2021understanding}, it is rarely done by a single person.
\added{According to ISBSG~\cite{isbsg}, 64\% of software development teams have at least five members.}  
Human involvement has led stakeholders to acknowledge that software development is a social activity~\cite{dittrich2002social, rohde2017grounded, mothman}, where interpersonal interactions significantly influence project outcomes~\cite{humor}.
Therefore, the software development industry developed various strategies to improve project management processes~\cite{effectiveManagement, agileArt, mothman}.
\added{Project management processes facilitate interpersonal interactions by defining roles, establishing communication channels, and fostering collaboration. 
These processes play a vital role in aligning team efforts and positively influencing project outcomes.}

The research community has developed various strategies to address the impact of social interactions on software engineering.
One notable approach is the socio-technical framework first proposed by Cataldo et al.~\cite{cataldo2008socio}.
The framework's core concept is that the coordination requirements set by task dependencies should align with the actual coordination activities performed by engineers.
Additionally, researchers have introduced concepts to capture how poor decisions affect software development, giving name to unwanted impacts of social interactions.
Tamburri et al.~\cite{tamburri2013social} adopted the idea of \textit{social debt} from social studies~\cite{muir1962social}.
In simple terms, social debt is the additional cost occurring when strained social and organisational interactions affect smooth software development~\cite{tamburri2015social}. 
Tamburri et al.~\cite{tamburri2013social} then focused on the aspects of team-based software development to identify the sources of social debt, naming the corresponding processes \textit{community smells}.
This approach allowed researchers to capture issues within the software development framework. 
Palomba et al.~\cite{palomba2018beyond} established a connection between community smells and code smells, and suggested approaches to address the underlying issues.

However, the community smell approach does not capture every problem in teamwork. 
For instance, the code review ping-pong~\cite{dougan2022towards} is a teamwork problem that has never been studied as a community smell~\cite{espinosa2021understanding}.
This particular issue falls into the code review domain and thus was identified in code review studies.
It remains possible that certain teamwork problems and associated malpractices cannot be easily assigned to any domain and thus remain invisible to the research community.
% When studies focus on specific parts of software engineering activities or try to tackle problems that fit into predefined frameworks, certain issues might remain invisible to the research community.
One approach to discover these problems is to explore an open-ended question:
\observation{\centering
\textit{What could possibly go wrong in software engineering teamwork?}}

Once a team identifies a teamwork related problem, they can try different strategies to manage it. 
The options range from the adoption of automated management tools to the implementation of organization-wide strategies~\cite{espinosa2021understanding, okr}. 
Large-scale solutions like organization-wide strategies can address the underlying socio-technical issues. 
\added{For example, a project Integra~\cite{tamburri2019software} described by Tamburri uses Architectural Board of people who are responsible for making architectural decisions.
The board addresses socio-technical issues which lead to the architectural community smells such as Obfuscated Architecting.
To achieve this, the board meets biweekly in order to 1) make and spread architecture decisions and 2) analyze how organization culture affects creating and spreading decisions.}

However, the organization-wide strategies often come with associated switching costs and may encounter resistance from employees~\cite{kim2010effects}.
These switching costs may even outweigh the benefits for the team or organization, prohibiting the desired change. 
For example, Jabrayilzade et al.~\cite{jabrayilzade2022bus} find that the lack of resources within a team is one of the reasons why the bus factor problem often remains unaddressed.
Hence, addressing a teamwork problem with a tool that demands minimal time investment from employees could be more pragmatic than attempting a complete resolution through a company-wide policy.
\added{For example, a company can address ``Missing context in reviews'' code review smell~\cite{dougan2022towards} by implementing a policy that requires engineers to write descriptive commit messages.
Another option could be adopting a deep learning-based tool that suggests commit messages to engineers.
While the auto-generated commit messages may not be perfect, they would likely yield improvements in some cases and potentially face less backlash.}

This paper aims to study undesirable patterns in collective development, including identifying the patterns, understanding their characteristics \added{(frequency, signigicance, consequences)}, and evaluating their addressability by tools.
An \textbf{\up in collective development}, or \textbf{\up} for short refers to any teamwork problem or practice that might not currently be problematic, but has the potential to evolve into a problem if left unattended.
\added{Further in this paper, the word "pattern" is also used as a short version of "\up".}

\added{\textit{What is the difference between an undesirable pattern and the notion of a community smell?}}

\added{An \up encompasses a broad range of collective development challenges, including both social and technical aspects.
A community smell refers to a set of suboptimal organizational structures that lead to the emergence and accumulation of both social and technical debt.
While ``\up'' is a broader term encompassing various potential teamwork problems, ``community smell'' is a more specific concept highlighting organizational inefficiencies within the organizational framework.}
\added{Finally, in this paper we use the word "address" with the meaning of mitigating a problem or a part of a problem which might make practitioners' lives easier.}

\subsection{Research Questions and Study Overview}\label{sec:intro_rq}
In this work, we aim to answer following research questions:
\observation{
\begin{itemize}
    \item{\textbf{RQ1}} What teamwork-related \ups do IT practitioners face in their work?
    \item{\textbf{RQ2}} What is the relative significance and frequency of these \ups?
    \item{\textbf{RQ3}} What tools \added{or features for existing tools} might help practitioners address teamwork related \ups?
\end{itemize}
}

We conduct a qualitative and a quantitative study to answer these research questions.
In the first phase of the study, we conduct \ninters in-depth exploratory interviews with various \added{IT practitioners} from 17 different countries and work experience ranging from 1 to 26 years.
As a result, we identify a list of \nprobs \ups (RQ1).
We also ask the interviewees to imagine hypothetical tools or features that could help address these patterns.

Using the collected lists of tools and patterns, we conduct two surveys. 
The first survey, \surveyOne, with \nsurveyOne participants, assesses the relevance of the identified patterns and their importance in terms of significance and frequency (RQ2). 
In the second survey, \surveyTwo, IT practitioners evaluate the potential usefulness of the tools proposed by the interviewees in the previous phase of the study (RQ3). 
We also ask participants how much influence they have among their team members in regards to adopting new tools and practices. 
Roughly 600 out of \nsurveyTwo respondents were individuals with a significant influence on adopting new tools. 
We assume that their opinion on tools' viability is more correlated with the likelihood of tool adoption.
Therefore, the data on their opinions might allow decision-makers from the industry to make better informed decisions on tool development.

The contributions of this paper are:
\begin{itemize}
    \item We collect a list of teamwork-related \ups and classify them in terms of origins and consequences (\Cref{sec:results_interview}). \added{For ten \ups, we also gather a list of ideas of tools and features that might address the patterns.}
    \item Using the results of the \surveyOne survey (\Cref{sec:results_problems}), we assess the impact of the collected \ups.
    \item Using the results of the \surveyTwo survey (\Cref{sec:results_tools}), we evaluate ideas of tools and features for existing tools that could potentially manage identified patterns.
\end{itemize}
The results of this study can serve as a foundation for conducting more in-depth studies on specific \ups.
In addition, the results provide information on the potential value of future tools that help address the \ups in teams of practitioners.

The rest of the paper is organized as follows.
In \Cref{sec:background} we review the research on the teamwork problems and the corresponding approaches to their management.
In \Cref{sec:methodology} we outline the methodology of this study and link it to the research questions. 
% We explain how we identify \ups, and categorize these patterns and their consequences.
% We also explain, how we gather and evaluate tool and feature ideas for addressing these patterns.
Next, there are three pairs of sections, each related to one of the study phases: qualitative interviews (\Cref{sec:design_interview}, \Cref{sec:results_interview}), the \surveyOne survey (\Cref{sec:design_problems}, \Cref{sec:results_problems}), the \surveyTwo survey (\Cref{sec:design_tools}, \Cref{sec:results_tools}).
Each pair consists of a design section that describes the structure, demographics, and analysis processes, and a results section that presents the detailed outcomes and respective analyses.
In \Cref{sec:discussion}, we discuss the results of this study, its implication for research and industry, and the ideas for future research directions.
In \Cref{sec:validity}, we discuss potential threats to the validity of this research and address ethical concerns.
Finally, in \Cref{sec:conclusions} we summarize out findings and conclusions.

\section{Background}\label{sec:background}
\subsection{Problems in collective development}
Software development often involves collaboration.
Teamwork facilitates fruitful discussions, and simplifies weeding out bugs and suboptimal architectural solutions~\cite{van2009shallow}. 
However, it can also result in imperfect social interactions that hinder the team's ability to achieve its goals.
\added{For example, team work requires information sharing, yet it also introduces various risks, such as multitasking and a lack of continuity within the development team~\cite{ghobadi2017risks}.}

The problems in teamwork may have different manifestations and come from different sources.
In this subsection, we overview studies on problems in software engineering (SE).
We highlight how different groups of issues and antipatterns may be teamwork problems, even if they do not look like such.

\subsubsection{Community smells}
Tamburri et al.~\cite{tamburri2015social} were the first to introduce a notion of \textbf{community smell}.
The community smell is a set of suboptimal organizational structures that lead to the emergence and accumulation of both social and technical debt.
The social debt is additional project cost caused by sub-optimal social or socio-technical decisions.
As opposed to technical debt, this extra cost is “social" in nature, i.e. connected to people and their organisation~\cite{tamburri2015social}.
\added{This difference is in the form of the debt, not in its causes: the technical debt can also be caused by bad organizational or architecture decisions}~\cite{Tdebt1, Tdebt2, Tdebt3, Tdebt4}.
\added{Social debt can be accumulated when an IT company dismantles a team that worked on a legacy project without explicitly transferring project ownership.~\cite{tamburri2013social}.}
Unlike code smells, community smells cannot appear in a project developed by a single person.

Software engineering researchers have been actively studying community smells in the recent years~\cite{cataldo2008socio,catolino2020refactoring,catolino2021understanding,tamburri2015social,tamburri2013social,tamburri2019exploring}. 
These studies shed light on the properties of the community smells, some of which we cover in the rest of the subsection.
Our presentation of community smells often follows the systematic review by Caballero-Espinoza et al.~\cite{caballero2022community}.

The research on smells often focuses on automatic detection of the smells (see~\cite{codedetect1, codedetect2, codedetect3} \added{for code smells} and~\cite{almarimi2021csdetector, tamburri2019exploring, comdetect1} \added{for community smells}).
With tools that automatically detect community smells, researchers can check whether these smells are present in open source projects~\cite{almarimi2021csdetector}.
Moreover, IT practitioners can use automatic detection tools to check whether they have community smells in their projects.
Finally, these tools can be used to unearth the impact of the community smells on a project.
Palomba et al.~\cite{palomba2018beyond} use \texttt{CodeFace4Smells} tool to detect several community smells. 
For example, they detect the \textit{Lone Wolf} community smell, which corresponds to the group of engineers who collaborate on an artifact, but do not communicate with each other.
They then use the collected data to demonstrate that the presence of community smells influence the severity of code smells.

Not every community smell will manifest itself in the mineable data produced in the course of software development.
For example, Tamburri et al.~\cite{tamburri2016architect} identify the \textit{Priggish Members} community smell.
This smell corresponds to demanding pointlessly precise compliance with standards, rules, and guidelines from the community members.
If a Priggish Member only participates in meetings and chats, this smell will not appear in mineable data such as code reviews or issue trackers discussions.
Thus, some community smells can be hard to identify automatically. 
Moreover, they can be invisible to individuals external to the team or even to some team members.

The invisibility of community smells to team members is increased by their innocuousness~\cite{caballero2022community}.
A community smell is not a ``show-stopper'' per se~\cite{tamburri2019exploring}, but is rather a circumstance that will later increase the cost of development.
This can be contrasted with the notion of the \cdp, first suggested by Jabrayilzade et al.~\cite{jabrayilzade2022bus}.
A \cdp is a problem perceived as a factor that already hampers project development. 
Jabrayilzade et al. hypothesise that every \cdp is a result of aggravation of one or several community smells.
Moreover, some of the \cdps found by Jabrayilzade et al. do not have clearly matching community smells described in the literature.
Thus, the problems in collective development noticed by practitioners may be different from the community smells found by researchers.

While the literature distinguishes 30 different community smells, most of the studies focus on only four of them: \textit{Organizational Silo, Black Cloud, Lone Wolf, and Radio Silence/Bottleneck} ~\cite{caballero2022community}.
The majority of the community smells extracted by Caballero-Espinoza et al.~\cite{caballero2022community} only appear in one to several studies.
Moreover, according to the review, the causes or effects for some of the community smells like \textit{Code Red} are not described in the literature.
As researchers lack common approach for reporting causes and effects of community smells, it is hard to compare the impact of different smells.
All these factors can preclude practitioners from analyzing their projects through the lens of community smells.

\subsubsection{Teamwork problems which are not community smells}
Not every problem in collective development manifests itself as a community smell. 

\textbf{Code smells} in source code is a surface indication that usually corresponds to a deeper problem in the system~\cite{fowler}.
While the majority of code smells are not exclusive to teamwork, some of them might happen explicitly due to shortcomings in collaborative work. 
For example, the \textit{Duplicated code} code smell might arise if two team members independently implement the same functionality.
We use the recent exhaustive review of Jerzyk et al.~\cite{jerzyk2023code} as a reference for code smells known to the community.

\textbf{Code review smells} according to Do{\u g}an and T{\"u}z{\"u}n refer to bad practices within the code review process that can result in process debt~\cite{dougan2022towards}.
The process debt, correspondingly, is a sub-optimal activity or process that might have short-term benefits, but generates a negative impact in the medium-long term~\cite{martini2019technical}.
As code review is a team practice, all code review smells are related to teamwork problems. 
It is also possible that code review smells may be symptoms of deeper teamwork problems.

\subsection{Managing problems in collective development}
Caballero-Espinosa et al. identify various approaches for addressing community smells.
This line of thinking can also be extended to the management strategies of other problems in collective development.

\textit{Organizational strategies} are socio-technical decisions designed and implemented by the organizations.
For example, an organization can require engineers to perform code reviews in two working days after receiving a request, addressing the \textit{Sleeping reviews} code review smell.

\textit{Tools and Models} correspond to the approach of automated problem discovery and assessment.
They can come in three different kinds.
First, researchers created dedicated tools such as csDetector~\cite{almarimi2021csdetector} \added{or CodeFace4Smells~\cite{tamburri2019exploring},} which predict community smells using repository data.
These tools utilize social network analysis data, or use ML to identify particular types of problems. 
Unfortunately, the practitioners did not widely adopt these tools.
Second, there are dedicated business tools, such as Qodana~\cite{qodana} or SonarQube~\cite{sonarcube}, which provide insights into the quality, health, and maintainability of software codebases.
\added{For example, SonarQube~\cite{sonarcube} provides secrets detection to avoid inadvertent security leaks, which addresses one of the \ups our respondents identify in the interviews}, see~\cref{sec:results_interview}.
The goal of these tools is to address the practitioners' problems. 
\added{However, we hypothesize that the problems these tools address may not correspond directly to what researchers or practitioners perceive as a problem.}
Finally, there are teamwork tools such as Slack~\cite{slack} or Confluence~\cite{confluence}.
Although they were not originally designed to address collective development problems, they can still be applied for that purpose.
For example, writing proper documentation in Confluence can eradicate the \textit{Invisible architecting} community smell.

\textit{Frameworks} like SPACE~\cite{forsgren2021space} can be seen as a blend between organizational strategies and models. 
Such frameworks offer a set of metrics along with stability thresholds that predict the variability of problems. 

\textit{Guidelines} for addressing community smells consist of suggestions for managing team composition by considering characteristics of teams or communities~\cite{tamburri2016architect}, \added{and improving consistency of architectural decisions~\cite{wohlrab2019improving}.
The suggestions of these guidelines provide clear call to action that could address some of the community smells.
For example, guideline 6 by Wohlraub et al.~\cite{wohlrab2019improving} calls for using one accessible source of information, which could reduce risks of the Architecture by osmosis community smell.}

\observation{
In summary, while the community smells approach allows to study some teamwork problems, not all problems fit this mold. 
For example, teamwork problems related to the code smells and code review issues, may diverge from the community smell framework. 
This study embraces an open-ended approach to unearth and examine a diverse set of teamwork challenges, aiming for a comprehensive understanding beyond existing frameworks.
}

\section{Methodology}\label{sec:methodology}
In this section, we briefly review our study methodology, as applied to the research questions we pose.
We offer a more detailed description of the study design in \Cref{sec:design_interview}, \Cref{sec:design_problems}, and \Cref{sec:design_tools}.

\begin{figure}[htp]
\centering
\includegraphics[width=\textwidth]{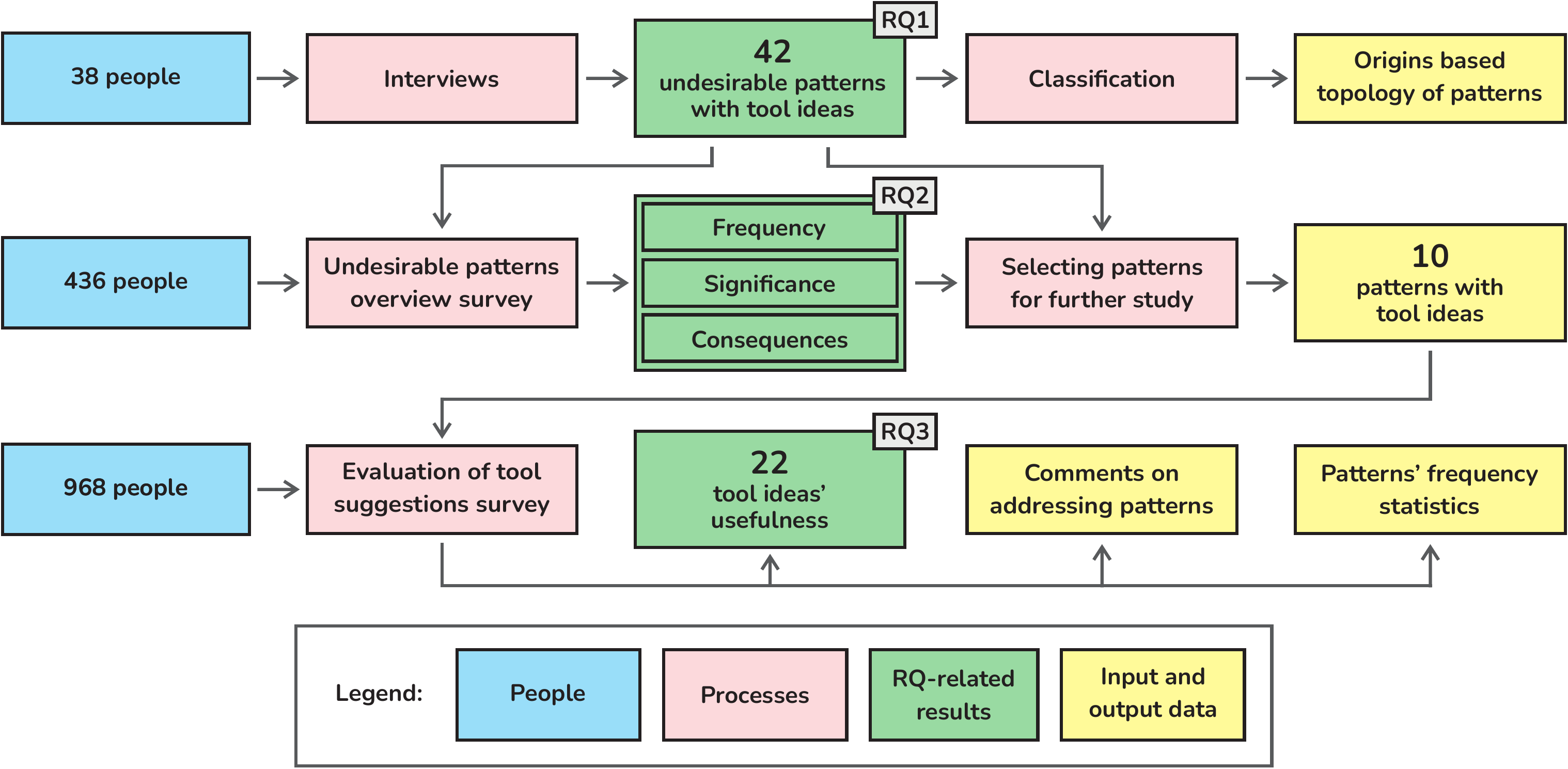}
\caption{Pipeline of the entire study.}
\label{fig:scheme_study}
\end{figure}

This study includes three phases:
\begin{enumerate}[topsep=0pt]
    \item In-depth exploratory interviews;
    \item The \surveyOne survey;
    \item The \surveyTwo survey.
\end{enumerate}
\Cref{fig:scheme_study} depicts the pipeline of the whole study.

In the first phase of this study, we conduct exploratory interviews to compile a comprehensive list of teamwork problems. 
This approach helps minimize the impact of our biases and preconceptions on identifying existing teamwork issues.

To recruit participants for our surveys and interviews, we reach out to individuals from the \jb\ mailing database.
This database contains e-mails of people who gave consent to \jb\ to receive marketing and analytics e-mails.
\added{While we cannot disclose the size of the database, we have sent more than ten thousand invitation e-mails for each of the surveys.}
We develop the interview script in collaboration with the \jb\ Market Research \& Analytics team.
Prior to the launch, both surveys underwent a thorough review by the \jb\ Localization and Copyediting team to ensure the removal of any inconsistencies or complexities.
To avoid any "first impression" bias, the surveys only include descriptions of \up, without the associated names that we present in this paper.

We choose to conduct two separate surveys, \surveyOne and \surveyTwo. 
Combining both surveys into one would result in a lengthy survey, that would take approximately 40 minutes to complete. 
Research by Deutskens et al.~\cite{deutskens2004response} indicates that longer surveys tend to have lower completion rates and a higher frequency of "do not know" responses. 
By splitting the surveys into two parts, we aim to make each survey more manageable in terms of time and complexity for the participants, potentially leading to higher response rates and more meaningful data.

\subsection{Teamwork-related \ups}
\textbf{(RQ1) What teamwork-related \ups do IT practitioners face in their work?}

To address RQ1 and identify the undesirable teamwork-related patterns faced by IT practitioners in their work, we conduct a series of semi-structured interviews. 
During these interviews, we ask IT practitioners to share their experiences regarding teamwork problems by using the following question:
\observation{
Could you recall the most recent situations where the problems arose in a team which would never happen for a single-person team?
}
With this wording, we intentionally depart from the existing notions related to collective development and use the most common language we could come up with.
We ask our respondents to describe the teamwork problems they encounter in as much detail as possible.
This approach reduces potential misunderstandings arising from differences in terminology.
We believe that this question is broad enough to capture most of the problems that can arise in teamwork, including community, code, and code review smells.
Therefore, we use the term \textbf{\ups in collective development}, or simply \textbf{\ups}.
This term encompasses any teamwork issue or practice that might not yet be a problem but, if left unaddressed, can become one.
In \Cref{sec:results_interview_problems} we match the \ups we identify with various smells described in the literature.

After analyzing the descriptions of various teamwork problems, we compile a list of \ups.
The descriptions provided by our respondents include detailed accounts of the preceding circumstances and the consequences of the problems they encountered.
These descriptions enable us to create a list of potential consequences associated with each \up.
Additionally, we utilize the accounts of prior circumstances to develop a tag-like system for categorizing the origins of these \ups.
We provide detailed description of the interview design in \Cref{sec:design_interview}.

\subsection{Frequency and impact of \ups}
\textbf{(RQ2) What is the relative significance and frequency of these \ups?}

To address RQ2 and assess the relative significance and frequency of various \ups, we conduct the \surveyOne survey.
In this survey, we ask respondents to evaluate the \ups that were identified in the previous interviews.
To gain a better understanding of the impact of each pattern, we also present respondents a predefined list of consequences of undesirable patterns.
We require the participants to select consequences that they have personally experienced.
Additionally, we provide an open option where respondents can add any other consequences that may not be listed. 
However, only around 5\% of the respondents utilize this open option.
We present these questions on frequency, significance, and consequences for individual \ups only if respondents indicate that they have encountered this particular \up in practice.
The results of this survey allow us to determine which \ups are more prevalent and detrimental from the perspective of IT practitioners.
We provide detailed description of the \surveyOne design in \Cref{sec:design_problems}.

\subsection{Tools that help address the \ups}
\textbf{(RQ3) What tools might help practitioners address teamwork related \ups?}

To address RQ3 and identify potential tools that could assist practitioners in dealing with teamwork-related \ups, we employ the following approach:
\begin{itemize}
    \item[1.] During the interview process, for every reported \up, we inquire respondents whether they believe a tool, whether existing or not, could help them manage the pattern they have experienced. 
    We encourage them to provide as many ideas as possible, with each idea serving as a feature or tool request. 
    This process results in the collection of descriptions for 52 different tools that address 32 distinct \ups. 
    \item[2.] We choose the ten patterns with the most viable tool proposals for further study.
    Evaluating tools for all 32 \ups in a single survey would be impractical due to sample size constraints. 
    \item[3.] In the \surveyTwo survey, we present respondents with five randomly selected \ups (out of the ten selected in the previous step).
    We inquire whether they have encountered these patterns in practice. 
    For each pattern they have experienced, we ask respondents whether the proposed tools would assist them in addressing the corresponding pattern.
    In addition to gathering general background information, we also ask respondents about the extent of their influence in adopting tools for their team. 
    If the respondents who make decisions on tool adoption have little interest in a tool, this tool is unlikely to be adopted.
\end{itemize}
We provide a detailed description of the \surveyTwo survey design in \Cref{sec:design_tools}.

\section{Interview design}\label{sec:design_interview}
\begin{figure}[htp]
\centering
\includegraphics[width=0.98\columnwidth]{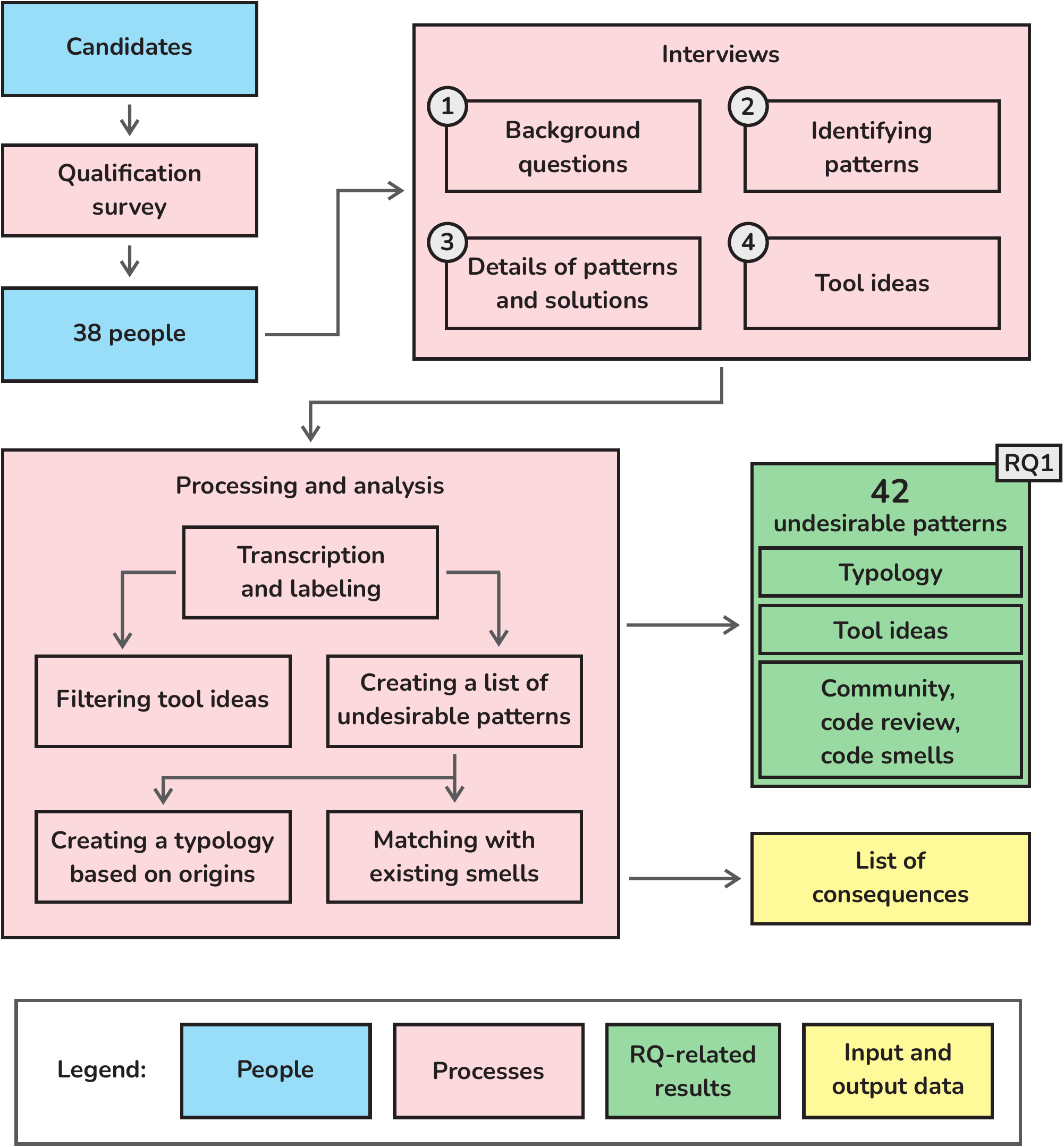}
\caption{Pipeline of the interview phase.}
\label{fig:scheme_interview}
\end{figure}

We conduct \ninters exploratory interviews to understand what \ups developers encounter in practice and how to address these patterns.
In total, we identify \nprobs different \ups.
We also ask the interviewees about the circumstances that contributed to the emergence of these patterns, their frequency, and the consequences they observed or heard of from their colleagues.
Based on this information, we create two surveys: \surveyOne and \surveyTwo.
These surveys allow us to collect additional data that significantly contribute to the statistical analysis, bolstering the validity of our findings.
\Cref{fig:scheme_interview} depicts the pipeline of the interview phase of the study.
The interview script is accessible in the Appendix B\footnote{Online Appendix: \url{https://github.com/KatyaKos/Undesirable-Patterns/blob/main/Undesirable_Patterns_Appendix.pdf}}.

\subsection{Recruitment and demographics}

\begin{figure}[htp]
\centering
\subfloat[Specialization (multiple choice question)]{
\includegraphics[width=0.48\columnwidth]{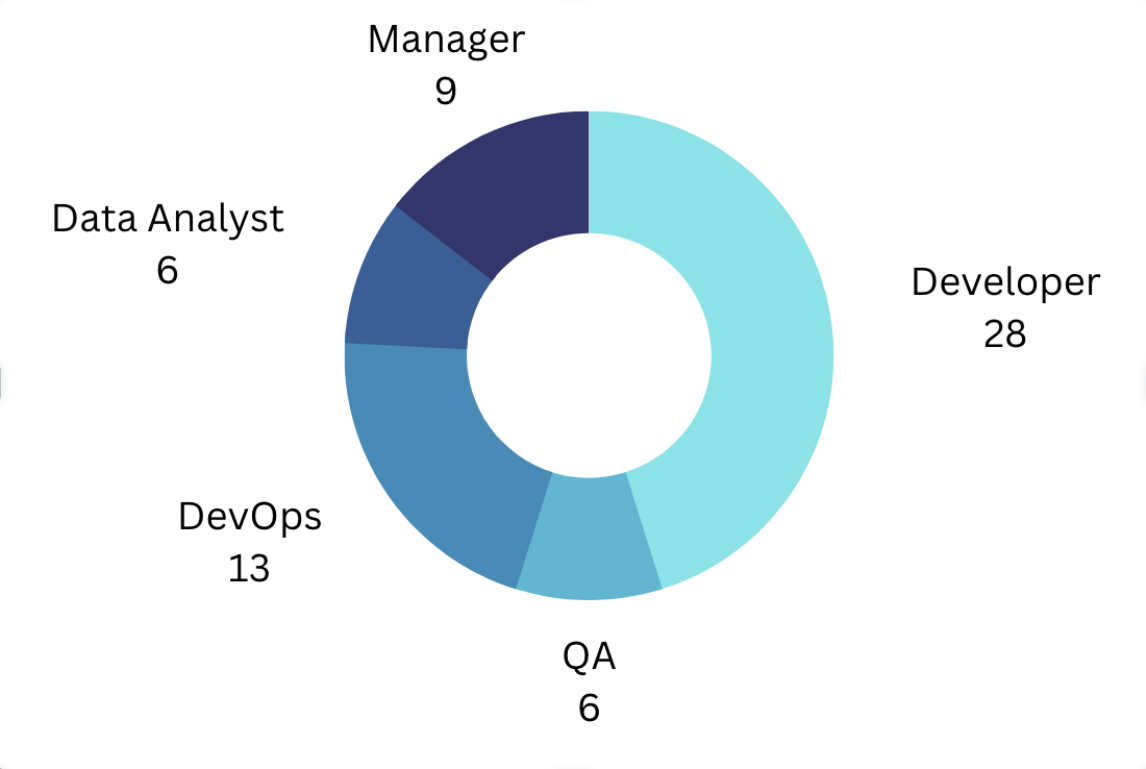}
\label{fig:int_dem_role}}
\subfloat[Team type (multiple choice question)]{
\includegraphics[width=0.48\columnwidth]{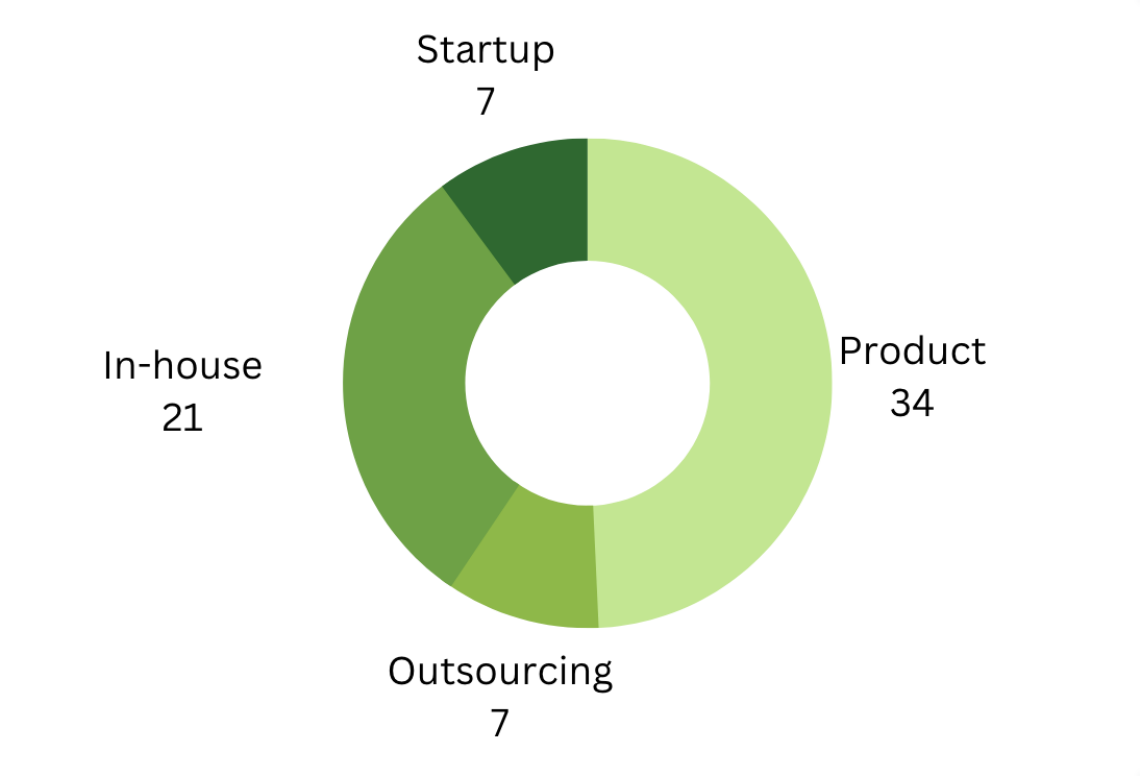}
\label{fig:int_dem_team}}
\\
\subfloat[Years of experience]{
\includegraphics[width=0.48\columnwidth]{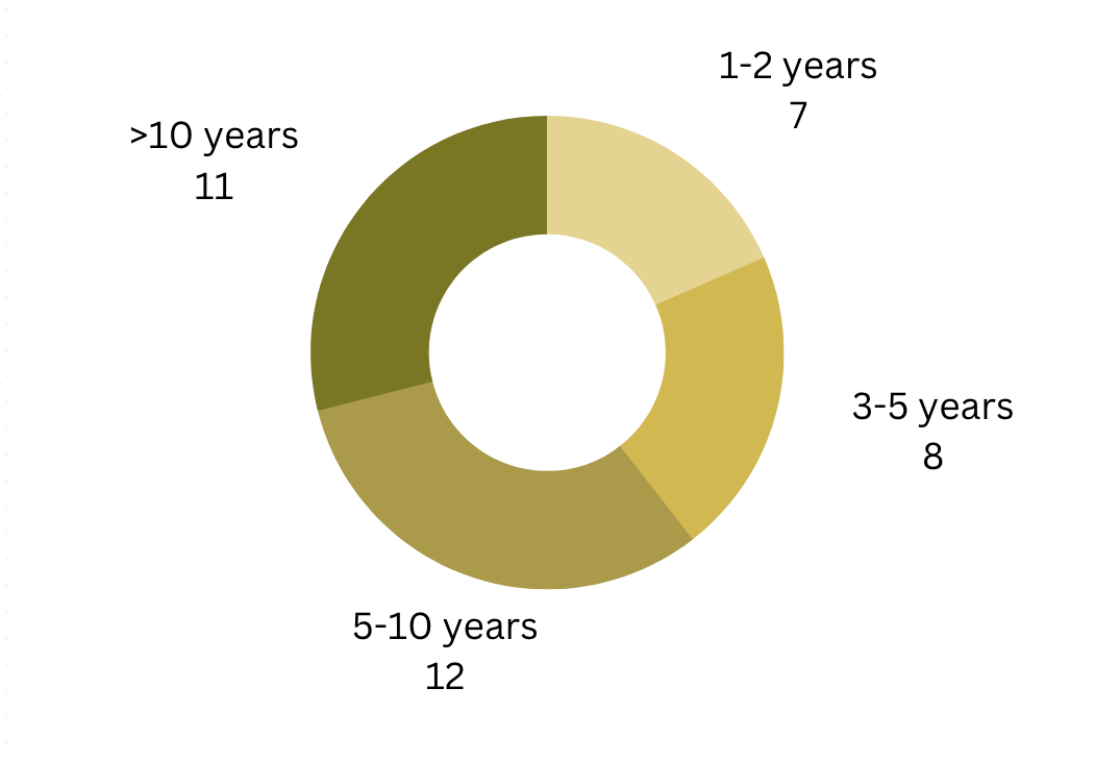}
\label{fig:int_dem_exp}}
\subfloat[Country of residence]{
\includegraphics[width=0.48\columnwidth]{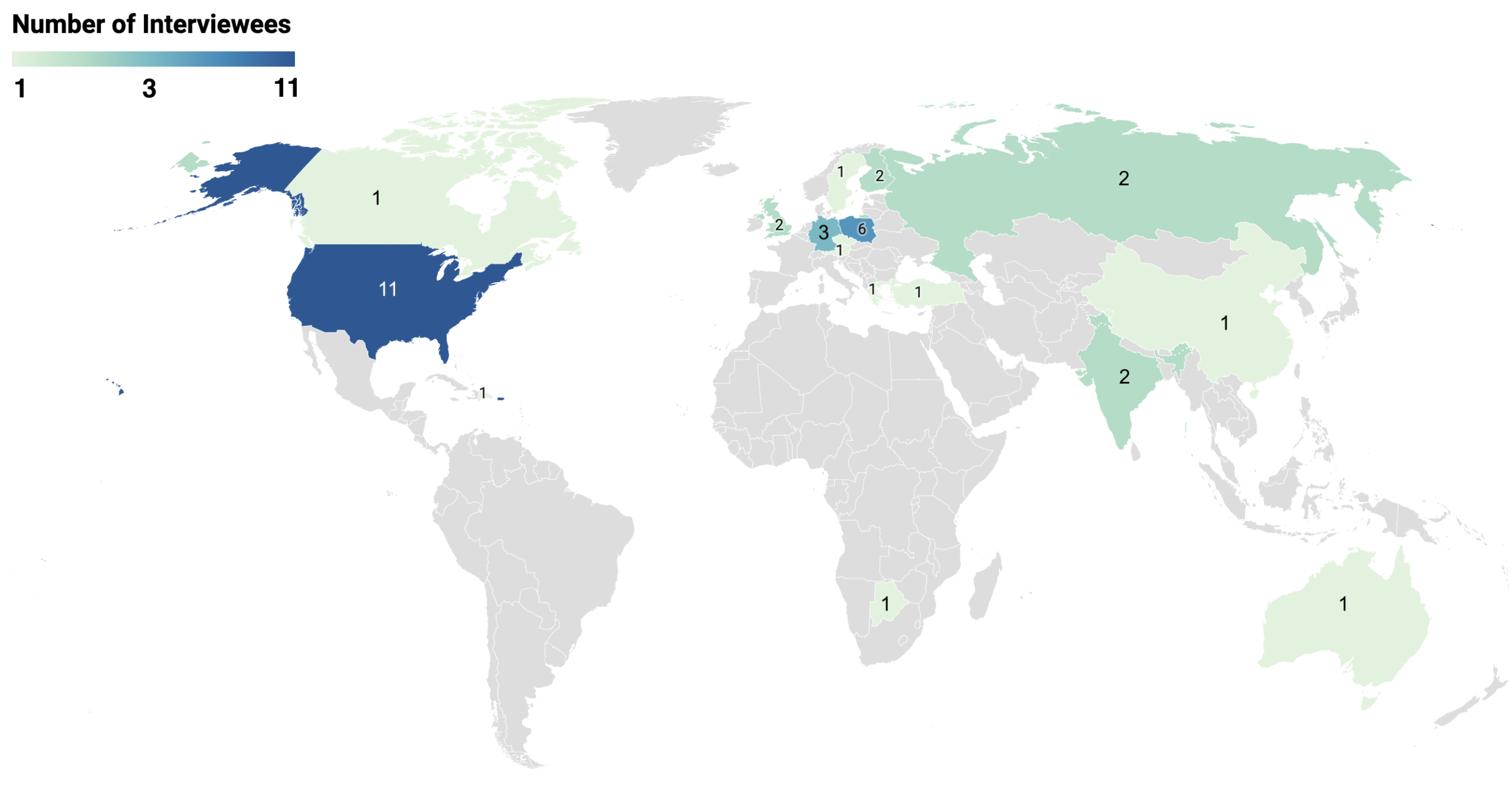}
\label{fig:int_dem_map}}
\caption{Demographic infographics for \ninters\ interviews.}
\label{fig:int_dem}
\end{figure}

For our interviews, we recruit \ninters practitioners with diverse backgrounds and varying levels of experience, representing a wide range of companies (\Cref{fig:int_dem}). 
All interviews are conducted in English by fluent non-native speakers.
Before being invited for an interview, each participant is asked to complete a brief qualification survey.
With this survey, we filter out unsuitable candidates and ensure a diverse range of respondents.

Interview participants reside in 17 different countries (\Cref{fig:int_dem_map}).
We only invite candidates who identify themselves as Software Developer, QA, DevOps, Data Analyst, or Manager (including Product Managers, Team Leads, CEOs, etc.)  (\Cref{fig:int_dem_role}).
We exclude candidates who do not work in IT industry (for example, retired).
We use the list of occupations which \jb extensively uses in the market studies.
Additionally, we classify their experience into four different buckets: 1 to 2 years, 3 to 5 years, 6 to 10 years, and more than 10 years (\Cref{fig:int_dem_exp}).
The participants had 1 to 26 years of work experience. 
Moreover, we broadly classify the software engineering teams into four categories based on the type of development they engaged in: Product teams, Outsourcing teams, In-house tool development teams, and Startup teams (\Cref{fig:int_dem_team}).

We recruit participants to ensure a minimum representation of five individuals within each category: role in development, type of development, and years of experience.
Notably, some interviewees fit under multiple labels within the same category, e.g. Team Lead and Developer, or Product team and Startup.
We provide our participants a comprehensive overview of the interview process through the qualification survey as well as prior to the interview itself. 
They explicitly consent to take part in the interview and agreed to have it recorded. 
As a token of appreciation for their participation, we offer interviewees a choice between a \$100 Amazon gift card or a one-year \jb All Product subscription as compensation.

\subsection{Interview outline and study protocol}
We conduct remote semi-structured interviews and record them for analysis. 
Initially, we ask participants general background questions related to their overall work experience as IT professionals, the team structures they work in, and the overall direction of their work \added{(e.g. frontend for a web shop or machine learning engineer for a bank).} 
Subsequently, we focus on exploring \ups witnessed by participants.
We place specific emphasis on patterns originating from imperfect interactions within their teams or organizations.
We intentionally exclude factors external to their organizations, such as the presence of toxic clients, from the scope of our investigation.

We employ a 1--5 Likert scale to assess the significance and frequency of the reported \ups. 
On the frequency scale, a rating of 1 indicates that the respondent has never encountered the pattern, while a rating of 5 indicates encountering the pattern in every or nearly every project. 
We provide references for all interim values.
Similarly, on the significance scale, a rating of 1 represents patterns with minimal impact on the project, while a rating of 5 indicates patterns that lead to project closures. 
We also give respondents the option to provide a range instead of a single number on each scale.
Additionally, we request the respondents to identify the most significant pattern in terms of impact and the most frequent pattern.

In the subsequent phase of the interview, we delve deeper into the reported \ups, initiating a comprehensive discussion. 
For each issue under consideration, we encourage the respondents to share a case story illustrating their personal experience with it. 
We ask the respondents to share what steps were taken to address the issue, what impact it caused and what measures were considered or implemented to prevent its recurrence.

At the end of the interview, we discuss with the interviewee the utilization and design of various tools aimed at mitigating the \ups they encounter. 
We ask the participants about their experience with using diverse tools to enhance team processes.
After that, we shift the focus towards exploring the challenges they face in their work from a tool-oriented perspective. 
For each pattern discussed, we ask respondents about their knowledge of existing tools or features thereof that could address the respective issues.
We also ask respondents to deliberate on the advantages and disadvantages of these tools. 
Furthermore, we encourage the participants to envision hypothetical tools or features that could potentially alleviate these problems.
These ideas could range from abstract concepts (e.g. 'An AI tool for automated documentation generation') to more specific ideas (e.g. 'An issue tracker feature that provides task time limits and requirements based on past tickets'). 
Additionally, we ask the respondents if there were any hypothetical tools or features unrelated to the discussed \ups that they would like to utilize in their work.

The interview sessions lasts between 60 and 90 minutes, with two interviewers (the first two authors of this paper) independently conducting each session. 
At the outset of the study, a member of the \jb Market Research \& Analytics team provides assistance during three interviews.

\subsection{Processing and analysis}
We utilize the \texttt{otter.ai}~\cite{otter} transcription tool to transcribe the audio recordings of the interviews.
The processing of the transcriptions involves collaborative participation from the 1st and 2nd authors. 
Initially, we process the general background information, as shown in \Cref{fig:int_dem}.
Subsequently, we assign short descriptions to the reported \ups.
For instance, a pattern such as ``My colleagues never write documentation and I often don't understand their code without it" becomes ``no documentation, can't understand code."
For each pattern, we include a short quote provided by the respondents to clarify the description, along with information on its frequency and significance according to the respondents. 
Additionally, we collect the consequences associated with each \up and derive their definitions from the descriptions provided by interviewees.
Finally, we gather ideas for tools and features that could potentially address the mentioned \ups.
\added{However, we filter out ideas that are too general or nonsensical, e.g. "AI tool that fixes it" or "a feature that blocks your office ID badge if there are any pending pull requests".}

During the subsequent phase of analysis, the 1st and 2nd authors compile the final list of reported \ups.
For example, the pattern of \emph{Insufficient Documentation} encompasses scenarios such as "My colleagues never write documentation and I often don't understand their code without it" and "I had to refactor old code but the documentation there was incredibly outdated, so it described absolutely different functionality." 
We also classify the identified patterns into broader categories based on their origins, e.g. \textit{Technical code-related} group and \textit{Social undesirable} group.
The two first authors independently develop the classification system, after which they collaborate to unify the identified categories.
Subsequently, both authors individually assign categories to each of the \ups and engage in a series of discussions to resolve any disagreements that arise during the process.

Additionally, we conduct a comparative analysis by matching the identified patterns with the community smells presented by Caballero-Espinoza et al.~\cite{caballero2022community}, code review smells by Dogan et al.~\cite{dougan2022towards}, and code smells by Jerzyk et al.~\cite{jerzyk2023code}.
Certain \ups combine multiple smells.
\added{By associating a pattern with a smell, we indicate that it relates to this smell, and do not claim an one-to-one match.}
For example, the pattern \textit{Dissensus of opinions and approaches} is related to community smells \textit{Dissensus}, \textit{Organizational skirmis}, and \textit{Solution defiance} by Caballero-Espinoza et al.
Additionally, not all smells have an analogue among the identified \ups, e.g. community smells "Unlearning" and "Prima donnas".

\observation{
Ultimately, we obtain a comprehensive list of \nprobs \ups, each accompanied by the following information:
\begin{itemize}
    \item Description
    \item Classification based on origins
    \item If possible, matching community, code review or code smells
    \item Up to three tool ideas proposed by the interviewees for addressing the pattern.
\end{itemize}
We also gather a list of consequences of these patterns that the interviewees experienced in their practice.
}
\section{Results and Conclusions from Interviews}\label{sec:results_interview}
Following the completion of \ninters exploratory interviews, we identify \nprobs distinct \ups \added{and tool and feature ideas offered for some of them.}
In this section, we define these patterns and introduce a typology system for the patterns based on their origin.
For detailed descriptions of the tools and their evaluation analysis, please refer to the results of the \surveyTwo survey in~\Cref{sec:results_tools_ideas}.

\subsection{Typology of the \ups}\label{sec:results_interview_typology}
The typology we present aims to clarify the origins of various \ups.
Each pattern belongs to one or two categories, and two patterns that share the same category can be argued to be more closely related to each other than two patterns without any common categories.

We identify three groups of the \ups based on their origin, with several subgroups per each group.
\begin{itemize}
    \item \textbf{Technical} patterns are recurring issues directly tied to some attribute, e.g. code, documentation, databases, or tools.
    \begin{itemize}
        \item \textbf{Technical code-related} \ups originate from collaboration on code and imperfect code-related practices.
        \item \textbf{Technical tool-related} \ups originate from inefficient application of engineering tools (e.g. IDE, issue tracker, video communication and messenger services).
        \item \textbf{Technical information-related} \ups originate from imperfect usage and distribution of data and information resources (e.g. database, documentation, team knowledge base).
    \end{itemize}
    \item \textbf{Social} patterns involve interpersonal challenges among team members, like disagreements or communication problems, that arise at the individual level, usually within the team.
    \begin{itemize}
        \item \textbf{Social variance} \ups originate from differences in employees' prior experiences, such as cultural, educational, and work ethics backgrounds.
        \item \textbf{Social communication} \ups originate from imperfect social interaction practices between people (e.g. lack of discussions, incomplete feedback) that occur through conversations, messaging, or other communication channels.
        \item \textbf{Social \added{reluctance}} \ups originate from tasks and duties that engineers are reluctant to fulfill due to perceived boredom, lack of importance, or unattractiveness.
    \end{itemize}
    \item \textbf{Organizational} patterns arise from management or company-wide decisions, affecting overall coordination and planning, like poor management choices or suboptimal team coordination.
    \begin{itemize}
        \item \textbf{Organizational planning} \ups originate from imperfect work planning that fails to account for all details and risks, and may involve inaccurate estimates of resources such as time or money.
        \item \textbf{Organizational management} \ups originate from poor management decisions and approaches to leading the team or the project.
        \item \textbf{Organizational synchronization} \ups originate from suboptimal coordination of inter- or intra-team tasks and schedules within the company.
    \end{itemize}
\end{itemize}
We assign each category exclusively to patterns that \textbf{consistently} originate from the related source, rather than as a potential reason.
For instance, \textit{Goal misunderstanding} is a problem on a personal level related to communication, not always a management-level issue.
Thus, we categorize it under \textit{Social communication}. 
Another example includes \textit{High level code style disagreement} and \textit{Low level code style disagreement}.
These patterns are consistently related to code but can be caused by both social reasons (e.g., developers were taught different code styles) and organizational reasons (e.g., a team manager did not establish code style rules)."

\subsection{Identified \ups}\label{sec:results_interview_problems}
\begin{figure}[htp]
\centering
\includegraphics[width=\textwidth]{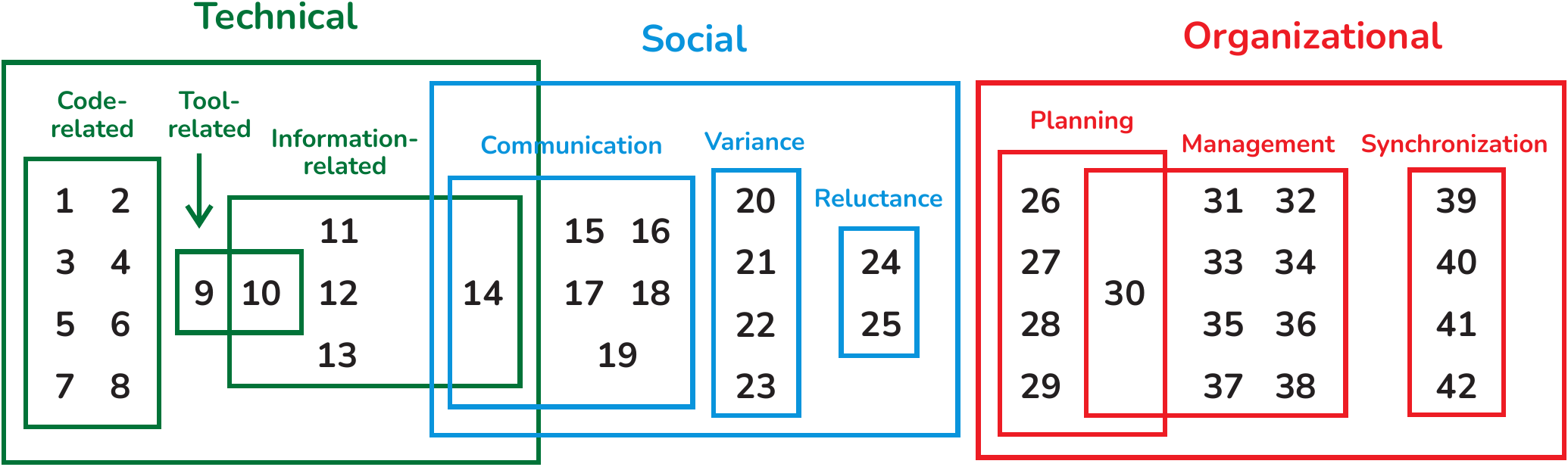}
\caption{Overview of \ups categorized by their origins.}
\label{fig:up_overview}
\end{figure}

In this section, we present a comprehensive list of the \nprobs identified \ups. 
Each pattern is accompanied by a descriptive name, definition, a typology based on the pattern's origins, and, if possible, matching smells presented by Caballero-Espinoza~\cite{caballero2022community}, Dogan et al.~\cite{dougan2022towards}, Jerzyk et al.~\cite{jerzyk2023code}.
\added{When an \up is associated with a smell, it indicates that the pattern relates to this smell, and do not claim an one-to-one match.}
We list technical patterns first, followed by social patterns, and conclude with organizational patterns.
We place patterns that fit into multiple categories accordingly between their related groups.
\begin{enumerate}
    \item \textbf{High level code style disagreement}. \\
    \textit{Categories: Technical code-related}\\
    %\textit{Jerzyk: Multiple}\footnote{There are many ``pairs'' of code smells which correspond to the extremes in software design, for example Middle Man and Message Chain. The code style disagreement in part may correspond to disagreement on what is the ``golden mean'' between these extremes.}\\
    A code style disagreement between colleagues which affects software design (e.g., OOP vs. functional, abstraction level, but NOT tabs vs. spaces). 
    The problem often comes up in code reviews.
    \item \textbf{Duplicated work}. \\
    \textit{Categories: Technical code-related}\\
    \textit{Jerzyk: Duplicate code}\\
    The same feature or similar piece of code was developed several times independently by different developers.
    \item \textbf{Low level code style disagreement}. \\
    \textit{Categories: Technical code-related}\\
    %\textit{Jerzyk: Fallacious method name, Uncommunicative name, Inconsistent names}\\
    Code style disagreement between colleagues on cosmetic issues, such as tabs vs. spaces, line length, or naming conventions (not the high-level abstractions such as OOP vs functional style).
    \item \textbf{Negligent code reviews}. \\
    \textit{Categories: Technical code-related}\\
    \textit{Dogan: LGTM reviews}\\
    Code reviewers do not put enough effort into reading the code and making comments on it, which might lead to bugs, improper code design, or an overcomplicated codebase.
    \item \textbf{\added{Uncoordinated code changes}}. \\
    \textit{Categories: Technical code-related}\\
    \textit{Caballero-Espinoza: Class cognition} \\
    Several people update the same code base simultaneously, without discussing the changes they made. 
    This could unintentionally introduce conflicts. 
    This may be particularly significant if the code base is large – Git rebase and editing will not work because of the high speed of new changes.
    \item \textbf{Slow code review process}. \\
    \textit{Categories: Technical code-related}\\
    \textit{Dogan: Sleeping reviews}\\
    Code review and pull request merging processes take too long due to slow responses from assignees.
    \item \textbf{Breaking change}. \\
    \textit{Categories: Technical code-related}\\
    An engineer tries to change code to meet their needs but inadvertently breaks some functionality used by other team members.
    \item \textbf{Outdated tests}. \\
    \textit{Categories: Technical code-related}\\
    Improper test maintenance. 
    This happens when tests are not updated, extended, or removed after the code base update.
    \item \textbf{Tool incompatibility}.\\
    \textit{Categories: Technical tool-related}\\
    Team members use different tools for the same goals.
    These tools might not work well together, causing workflow delay as engineers need to resolve compatibility issues. 
    This problem may affect both system components (such as libraries) and auxiliary tools (such as IDEs).
    \item \textbf{Multiple discussion channels}. \\
    \textit{Categories: Technical tool-related, Technical information-related}\\
    \textit{Caballero-Espinoza: Black cloud} \\
    Information is spread across several discussion channels used by team members (messengers, issue and task trackers, email, meetings, etc.), so it can be difficult for team members to understand exactly what is going on.
    \item \textbf{Insufficient documentation}.\\
    \textit{Categories: Technical information-related}\\
    \textit{Caballero-Espinoza: Invisible architecting} \\
    The existing documentation is not sufficient to understand the project because it is incomplete, badly written, very hard to understand, or was not updated for a long time.
    \item \textbf{Improper confidential data management}. \\
    \textit{Categories: Technical information-related}\\
    \textit{Caballero-Espinoza: Informality excess} \\
    Publishing passwords and private keys (e.g. by checking them in the repository).
    \item \textbf{Mutable data sharing}. \\
    \textit{Categories: Technical information-related}\\
    \textit{Jerzyk: Mutable data}\\
    Many team members use the same mutable data (e.g. the same database). 
    So, when one person modifies the data, the other team members may be unaware of the change and, when they use it, they will assume the data was not changed.
    \item \textbf{Faulty information sharing}. \\
    \textit{Categories: Technical information-related, Social communication}\\
    \textit{Caballero-Espinoza: Sharing villainy} \\
    Colleagues not being able to answer questions clearly, with full detail, or quickly enough.
    For example, when your colleague takes several days to answer a short question, or when they only give a short reply which is hard to understand without knowing all the related context.
    \item \textbf{Imperfect stakeholder communication}.\\ 
    \textit{Categories: Social communication}\\
    Stakeholders do not provide clear and complete data because either they cannot, do not want to, or do not know how to provide such data.
    \item \textbf{Long communication chain}. \\
    \textit{Categories: Social communication}\\
    \textit{Caballero-Espinoza: Leftover techie} \\
    Teams and/or stakeholders involved in a project do not communicate directly, so things deteriorate into a ``broken telephone" game. 
    For example, a customer communicates with the development team via sales managers and business analysts.
    \item \textbf{Hard to share task context}. \\
    \textit{Categories: Social communication}\\
    \textit{Caballero-Espinoza: Architecture by osmosis} \\
    \textit{Dogan: Missing context}\\
    Sometimes information cannot be easily transferred to another person.
    It might cause problems when the primary task assignee is planning to take time off work and needs to share all of their task related knowledge and data with other team members.
    \item \textbf{Goal misunderstanding}. \\
    \textit{Categories: Social communication}\\
    \textit{Caballero-Espinoza: Disengagement} \\
    Team members fail to understand the aim of a task, or a project, or a product, so the solutions do not align with the main goals. 
    For example, a team working on a product prototype does not create a minimal viable prototype, as expected, but instead works on a highly-functional scaling solution with a lot of features.
    \item \textbf{Lack of informal communication}. \\
    \textit{Categories: Social communication}\\
    Team members have no means of communicating informally.
    As a result, the team relationships become too official and engineers are not integrated into the wider company.
    \item \textbf{Passive engineer}. \\
    \textit{Categories: Social variance}\\
    A team member does not do anything unless they are specifically told to.
    \item \textbf{Cultural differences}. \\
    \textit{Categories: Social variance}\\
    \textit{Caballero-Espinoza: Cognitive distance} \\
    Diversity of the team members' cultural backgrounds or languages which causes unintentional misunderstanding or miscommunication between team members.
    \item \textbf{Dissensus of opinions and approaches}. \\
    \textit{Categories: Social variance}\\
    \textit{Caballero-Espinoza: Dissensus, Organizational skirmish, Solution defiance} \\
    The team cannot agree on how to resolve or approach a non-technical problem due to a difference of opinions or approaches. 
    For example, Alex wants to have three default reaction buttons for posts but Betty thinks it is better to have five buttons.
    \item \textbf{Technical dissensus}. \\
    \textit{Categories: Social variance}\\
    \textit{Caballero-Espinoza: Dissensus} \\
    Team members prefer different technical approaches and argue over which tool or language or framework to choose.
    \item \textbf{No task progress report}. \\
    \textit{Categories: Social reluctance}\\
    An engineer keeps working on a task without sharing their progress and issues with team members or managers. 
    This might lead to bugs or incorrect task execution due to a misunderstanding on the engineer’s part which, in turn, is also not communicated to their colleagues.
    \item \textbf{Ignoring directions}. \\
    \textit{Categories: Social reluctance}\\
    \textit{Caballero-Espinoza: Lone wolf} \\
    An engineer ignores the tasks or the directions given by the managers or technical leads. 
    This might lead to bugs, improper software design, or incorrect task execution, so the end result does not fit the requirements.
    \item \textbf{No acceptance criteria}. \\
    \textit{Categories: Organizational planning}\\
    There are no clear and well-described acceptance criteria (e.g. a task is formulated as “We need to speed up the loading process a bit”), making it impossible to understand what should be the end result of the task.
    \item \textbf{Insufficient planning}. \\
    \textit{Categories: Organizational planning}\\
    \textit{Caballero-Espinoza: Dispersion, Time warp} \\
    The team does not plan projects in enough detail. 
    For example, a frontend and a backend team do not plan out the API properly and design a product with incompatible parts.
    \item \textbf{Unclear requirements}. \\
    \textit{Categories: Organizational planning}\\
    No clear, well-thought-out, and formally defined requirements \added{for a task, a project, or a product} are given, so people interpret requirements individually.
    \item \textbf{Problem complexity mismanagement}. \\
    \textit{Categories: Organizational planning}\\
    A complicated problem has not been broken down into smaller, more manageable chunks for a developer to work on.
    \item \textbf{Disruption of plans}. \\
    \textit{Categories: Organizational planning, Organizational management}\\
    Work priorities change unexpectedly, so engineers are often being forced to work on something unplanned.
    \item \textbf{Poor onboarding process}. \\
    \textit{Categories: Organizational management}\\
    \textit{Caballero-Espinoza: Newbie freeriding} \\
    New engineers do not get enough work-related information to start working efficiently. 
    Often, the onboarding process contains only general information about office hours, benefits, vacation, etc. 
    However, engineers learn nothing about the team's code style or code review rules and do not get introduced to the current projects.
    \item \textbf{Not enough human resources}. \\
    \textit{Categories: Organizational management}\\
    The team’s workload is too high, so the team has to either work overtime or delay task implementation.
    \item \textbf{Inadequate leadership}. \\
    \textit{Categories: Organizational management}\\
    The group lead does not follow the best management practices (does not listen to team members, gives unethical feedback, constantly makes significant planning errors, etc.).
    \item \textbf{Toxic company culture}. \\
    \textit{Categories: Organizational management}\\
    Company culture makes team members uncomfortable, e.g. due to microaggression, bullying, or general animosity.
    \item \textbf{Underqualified colleagues}. \\
    \textit{Categories: Organizational management}\\
    \textit{Caballero-Espinoza: Obfuscated architecting} \\
    Having team members who are unable to properly solve a task or effectively communicate a problem due to lack of experience or knowledge.
    \item \textbf{Informational bottleneck }. \\
    \textit{Categories: Organizational management}\\
    \textit{Caballero-Espinoza: Code red} \\
    Bus/truck factor issue. 
    This happens when too few people know what is going on in some part of a project. 
    This might lead to a project stalling when those with the most knowledge become unavailable, because progress cannot be made on the relevant part of the project.
    \item \textbf{Insufficient financial resources}. \\
    \textit{Categories: Organizational management}\\
    A shortage in the team budget preventing the team from buying proper equipment, taking required business trips, or covering other necessary work-related expenses.
    \item \textbf{Unreasonable tasks}. \\
    \textit{Categories: Organizational management}\\
    The team receives tasks that are impossible to finish in the permitted time, tasks that contradict each other, or tasks that are otherwise unreasonable.
    \item \textbf{Synchronizing between teams}. \\
    \textit{Categories: Organizational synchronization}\\
    \textit{Caballero-Espinoza: Architecture hood, DevOps clash, Organizational silo} \\
    Different teams work on similar issues at the same time but cannot effectively synchronize their actions (meetings, statuses, rolling out into production, sharing information).
    \item \textbf{Teamwork pipeline bottleneck}. \\
    \textit{Categories: Organizational synchronization}\\
    A group of people cannot work efficiently together, because they cannot proceed without the results created by another group of people (e.g. the frontend team cannot develop without a part of the API being implemented by the backend team first). 
    The issue is not that one of the teams is too slow, but that the project development plan did not account for such bottlenecks.
    \item \textbf{Mismatching working hours}. \\
    \textit{Categories: Organizational synchronization}\\
    Different team members have different working hours and find it hard to collaborate due to little overlap in their working hours.
    \item \textbf{Inefficient meetings system}. \\
    \textit{Categories: Organizational synchronization}\\
    \textit{Caballero-Espinoza: Black cloud, Sharing villainy} \\
    Unreasonable meeting schedule (meetings are too long, too frequent, or too rare) which does not suit project needs.
\end{enumerate}

\observation{
    \added{
    We defined \nprobs \ups and linked some of them with the existing community, code, and code review smells.
    We also devised a typology system based on the origins of the patterns and used it categorize the \up list
    }
    (\Cref{fig:up_overview}).
}
\section{\surveyOne\ survey design}\label{sec:design_problems}
\begin{figure}[htp]
\centering
\includegraphics[width=0.98\columnwidth]{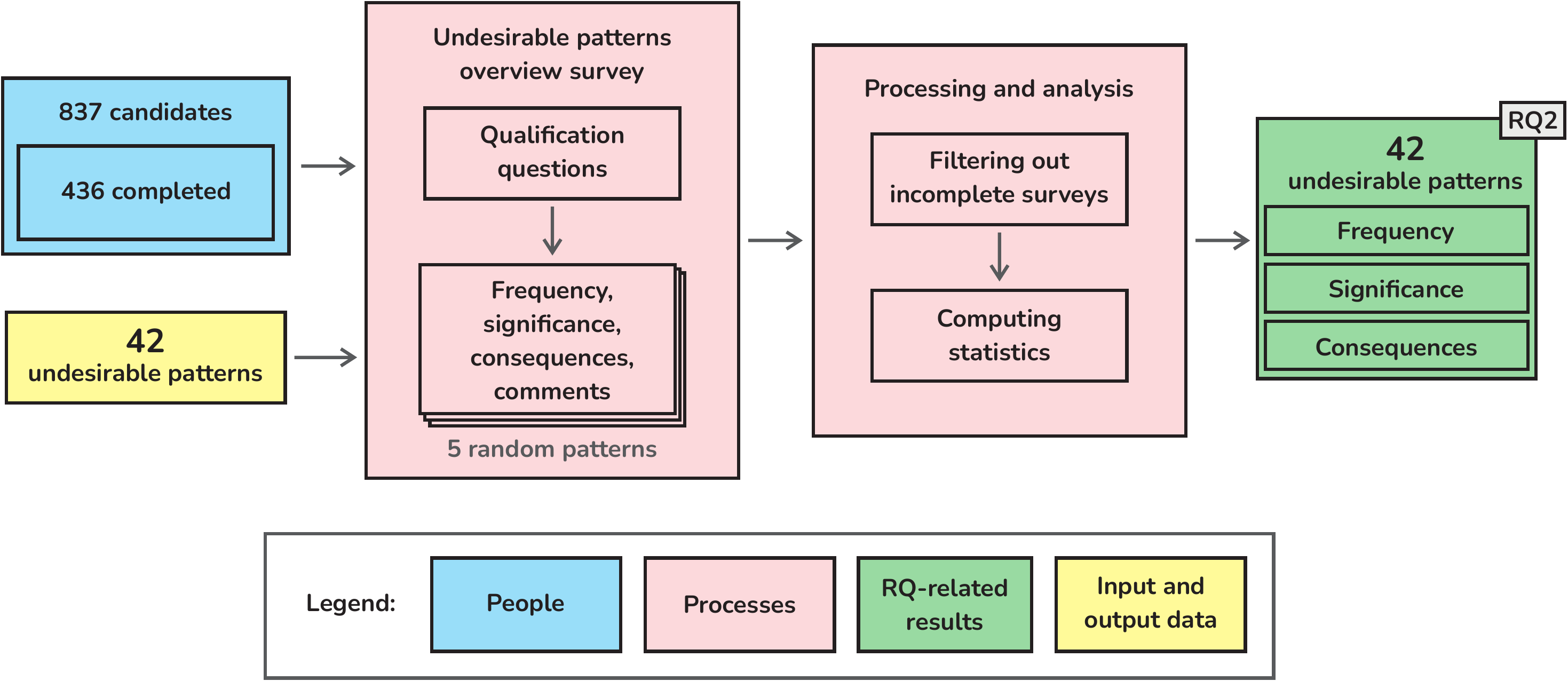}
\caption{Pipeline of the \surveyOne survey phase.}
\label{fig:scheme_surv1}
\end{figure}

The primary objective of the \surveyOne survey is to obtain insights from a larger pool of participants regarding the \nprobs identified \ups.
We ask the participants to provide their perspectives on the frequency and significance of these patterns in their own experience, as well as share the consequences they experienced throughout their professional careers.
\Cref{fig:scheme_surv1} depicts the pipeline of the \surveyOne survey phase of the study.
The survey script is accessible in the Appendix C\footnote{Online Appendix: \url{https://github.com/KatyaKos/Undesirable-Patterns/blob/main/Undesirable_Patterns_Appendix.pdf}}.

\subsection{Recruitment and demographics}
\begin{figure}[htp]
\centering
\subfloat[Specialization (multiple choice question)]{
\includegraphics[width=0.7\columnwidth]{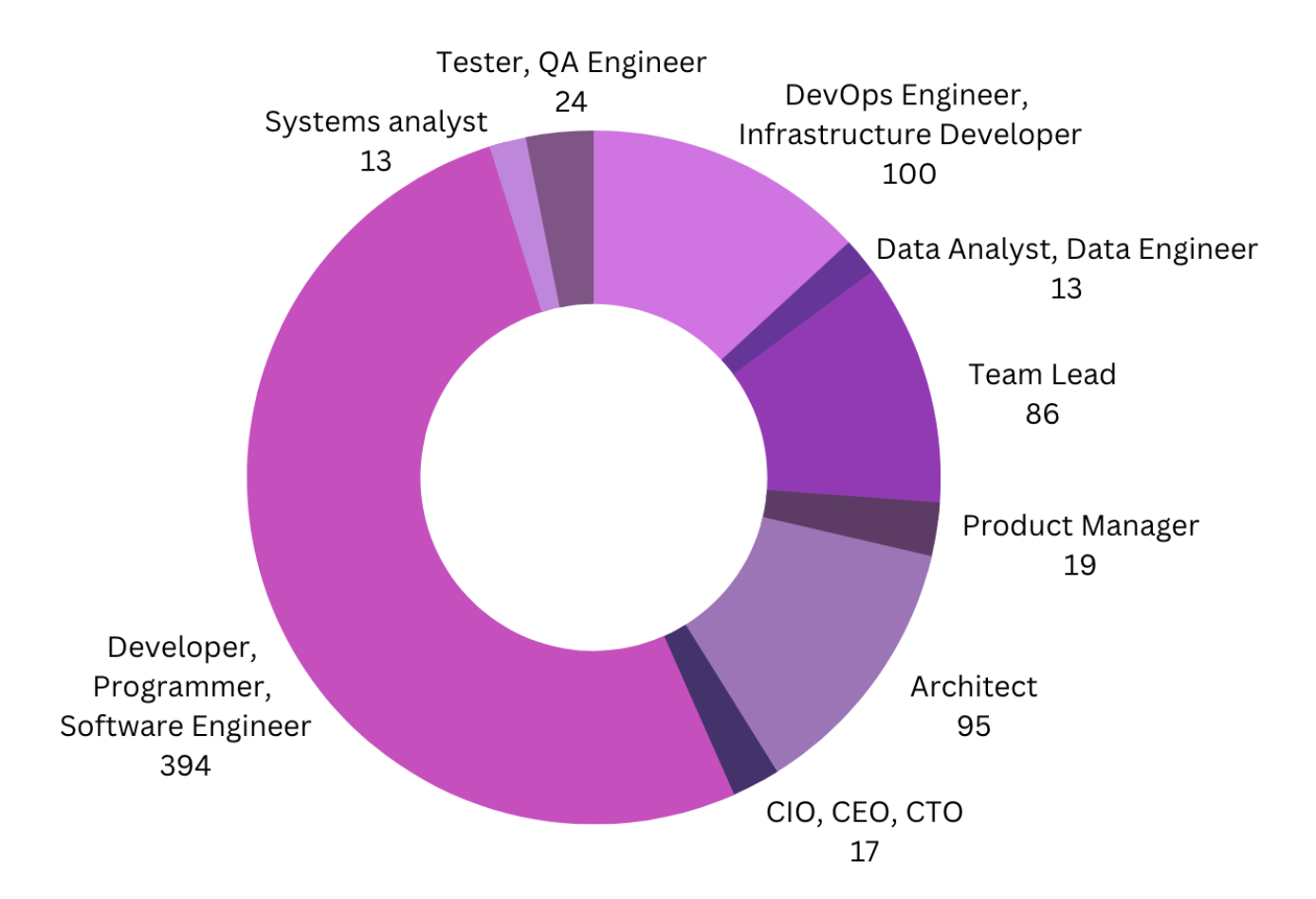}
\label{fig:surv1_dem_role}}
\\
\subfloat[Employment status]{
\includegraphics[width=0.4\columnwidth]{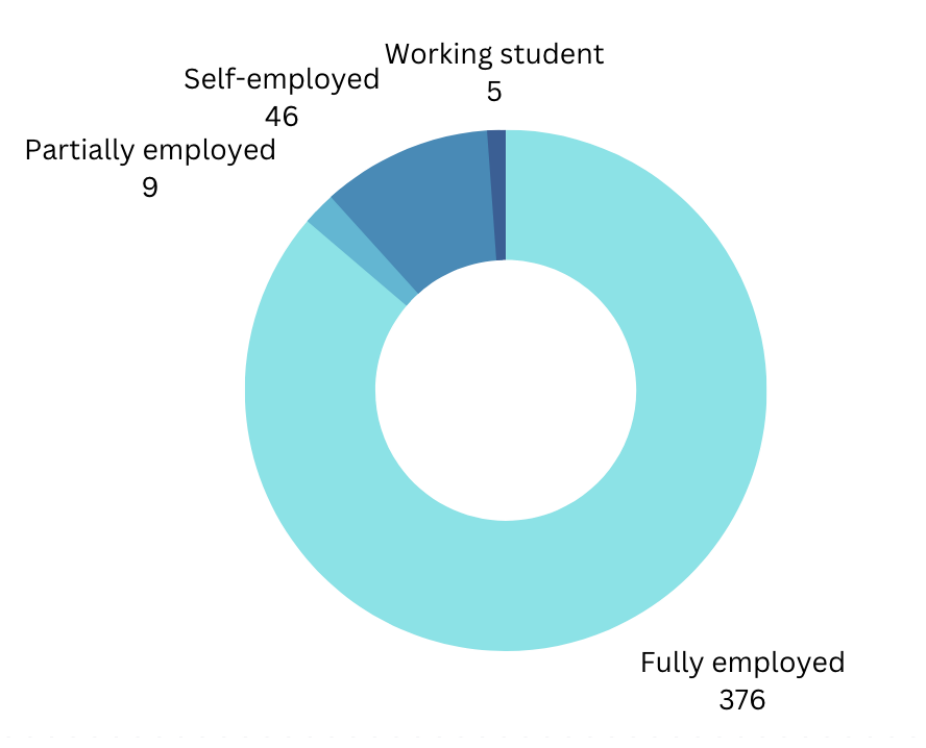}
\label{fig:surv1_dem_empl}}
\subfloat[Years of experience]{
\includegraphics[width=0.4\columnwidth]{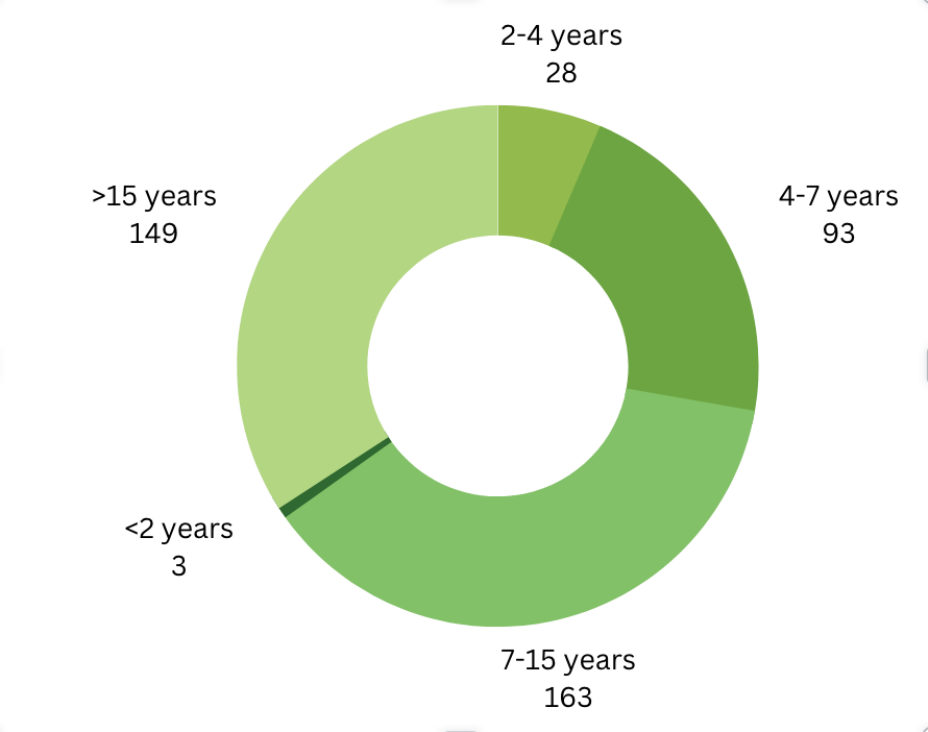}
\label{fig:surv1_dem_exp}}
\\
\subfloat[Participated in team projects]{
\includegraphics[width=0.4\columnwidth]{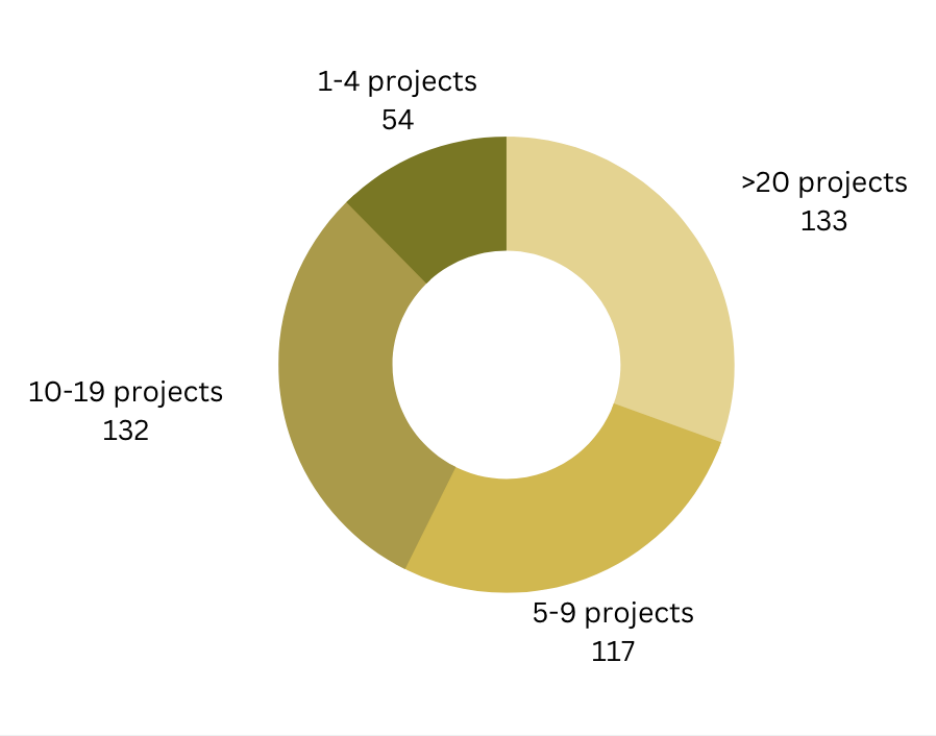}
\label{fig:surv1_dem_projs_part}}
\subfloat[Led team projects]{
\includegraphics[width=0.4\columnwidth]{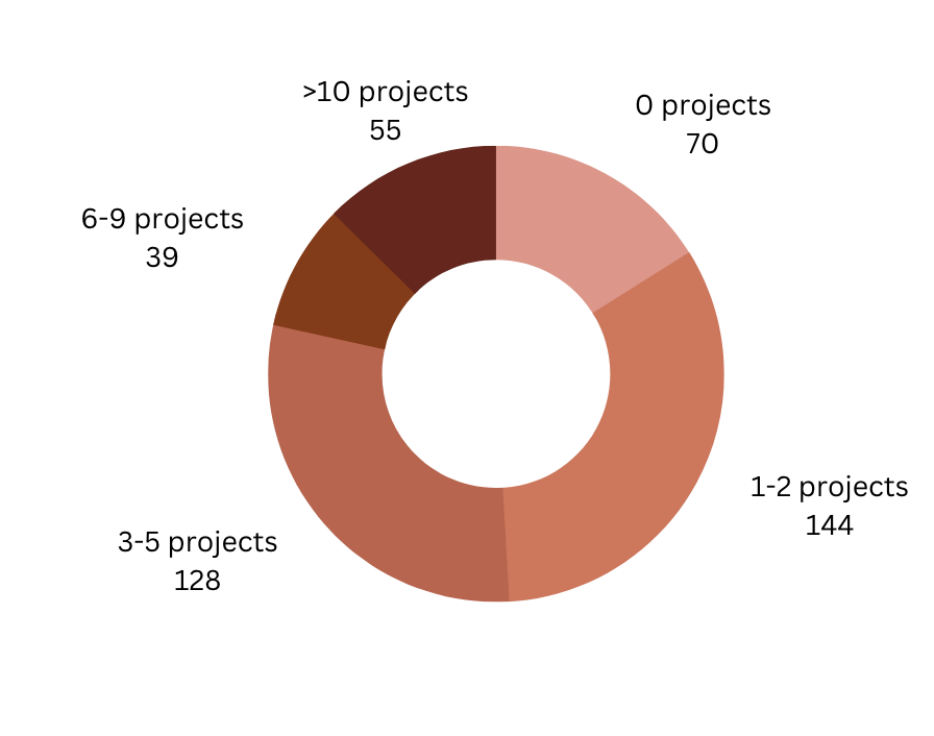}
\label{fig:surv1_dem_projs_led}}
\caption{Demographic infographics for the \surveyOne survey with \nsurveyOne participants.}
\label{fig:surv1_dem}
\end{figure}

For this survey, we recruit a total of \nsurveyOne practitioners with diverse backgrounds and varying levels of experience (\Cref{fig:surv1_dem}).
We conduct the survey in English.
The survey begins with the participants responding to questions about their background, and subsequently, unsuitable candidates are filtered out.
Initially, the survey attracted 837 candidates who answered the first question.
However, after the filtering stage, only \nsurveyOne individuals who met the criteria were eligible to proceed and ultimately completed the main survey.

The survey employs four filtering criteria.
The first stage involves an employment status filter, allowing only candidates who identify themselves as Fully employed, Partially employed, Self-employed, or Working students to proceed with the survey (\Cref{fig:surv1_dem_empl}). 
The second filter, \added{which excluded eight people, rejects} candidates who are not employed in the field of IT.
\Cref{fig:surv1_dem_exp} \added{displays the ratio of years of work experience reported by professionals who completed the survey.} 
In the third stage, we ask participants about their job roles, and the candidates who did not select any of the following roles are filtered out: CIO or CEO or CTO, Tester or QA Engineer, DevOps Engineer or Infrastructure Developer, Team Lead, Developer or Programmer or Software Engineer, Architect, Data Analyst or Data Engineer or Data Scientist, Product or Project Manager, System Analyst (\Cref{fig:surv1_dem_role}).
The final filter declines candidates who have not participated in any team projects throughout their careers (\Cref{fig:surv1_dem_projs_part}). 
Although we ask candidates for the number of projects they had led, we do not use this data for filtering (\Cref{fig:surv1_dem_projs_led}).

We inform the participants that the research findings will be shared with the community while ensuring the confidentiality of the information they provide. 
Furthermore, participants who successfully complete the survey are automatically entered into a raffle with a prize of their choice: a one year \jb All Products Pack subscription or a \$100 Amazon gift card.

\subsection{Survey outline and study protocol}
We conduct the survey using the Alchemer platform~\cite{alchemer}.
It begins with general background questions, which serve the purpose of filtering out unsuitable candidates.
Subsequently, we query the participants about the tools they regularly employ in their work.
Following this, the focus shifts towards exploring participants' experiences with \ups.
We make a clear distinction between "\ups" and "technical problems", with an explanation provided to ensure participants' understanding. 
\observation{
\textbf{Undesirable pattern} --- any teamwork problem or practice that might not currently be problematic, but has the potential to evolve into a problem if left unattended.
It can happen only to a team of two or more members.
\textbf{Technical problem} --- problem that can happen to any team, even a single-person one.
}
We then present the participants with five sections of questions, each dedicated to a randomly assigned pattern from the list of \nprobs patterns.
This ensures the random and uniform distribution of the \nprobs \ups among all participants.
On average, we received 51 responses for each pattern, with a maximum of 64 and a minimum of 39.

We will now outline the structure of each section, each section was related to one specific pattern.
The sections are similar, with the only variation being the name and description of the pattern in the beginning of the section.

Following the definition, we query the participants about the frequency of encountering the pattern within their experience. 
The available response options are as follows:
\begin{itemize}
    \item This problem is outside my domain of responsibility [Proceed to the next pattern]
    \item I've never encountered this problem [Proceed to the next pattern]
    \item I've encountered this problem, but it was an unusual situation
    \item I encounter this problem from time to time, but not very often
    \item I often encounter this problem
    \item Other (please specify):
\end{itemize}
The participants who indicate that the pattern is outside their domain of responsibility or that they have never encountered it were directed to the next section. 
We implement this approach to ensure that only participants with relevant experience would respond to subsequent questions regarding the pattern's significance and consequences.
Participants are also able to write their own answer in an open comment box.
The subsequent question evaluates the significance of the pattern, offering the following response options:
\begin{itemize}
    \item This problem hardly affects the workflow or does not affect it at all;
    \item This problem affects the workflow, but does not impact work-related processes (release deadlines are not moved, no overtime work is required, no need to involve managers or other outside people to address the issue). Intermediate goals, however, may have to be changed;
    \item This problem mildly impacts work-related processes (release deadlines are slightly moved, some features are postponed to be released later, managers participate in resolving the problem);
    \item This problem significantly impacts work-related processes (release deadlines are moved significantly as compared to the release cycle, some critical features are not shipped in time);
    \item This problem greatly impacts work-related processes (the project is closed or postponed indefinitely, some of the team members leave the team due to conflicts or general dissatisfaction).
\end{itemize}

To further explore the question of significance, we provide participants a list of consequences and ask to select the ones they have encountered when facing the pattern. 
Consequences can be divided into several broad groups that are not mutually exclusive.
Some of the consequences may be a corollary of the other; for example, massive departure of team members may result in project closure.
We group the consequences based on their primary impact.
Participants also have the opportunity to add any additional consequences not mentioned in the provided list. 
The comprehensive list of consequences includes:
\begin{itemize}
    \item \textbf{Wasting time} consequences cause deviation from the ideal timeline of the project development.
    \begin{itemize}
        \item \textit{Waiting} consequences correspond to the unnecessary delay in engineer's work caused by being dependent on others' input or actions.
        This may result in not having any work to do, or forgetting details and plans related to the task.
        \item \textit{Fixing or redoing} consequences correspond to the need to repeat or correct tasks.
        This may happen due to errors or deficiencies in the initial implementation or task description.
        \item \textit{Communication} consequences correspond to the inefficiencies in exchanging information, leading to time wasted in clarifications, or arguments.
        The communication can happen through meetings, calls, chats, or other means.
    \end{itemize}
    \item \textbf{Project development impact} consequences affect project lifecycle.
    \begin{itemize}
        \item \textit{Cancelled} consequences correspond to the project being terminated prematurely.
        This may happen due to budget constraints, shifting priorities, poor planning, or changing requirements.
        \item \textit{Insupportable} consequences correspond to the project becoming difficult or impossible to maintain, modify, or sustain over time.
        This may happen due to poor design or lack of documentation.
        \item \textit{Delayed} consequences correspond to the project falling behind schedule.
        This may result in disruptions to other project dependencies and increased over-time work, potentially leading to a decrease in work quality.
        \item \textit{Suboptimal or buggy} consequences correspond to the project being below the desired level of quality, efficiency, or functionality.
        This may happen due to errors, flaws, or performance issues.
    \end{itemize}
    \item \textbf{Team effects} consequences impact employees' work life.
    \begin{itemize}
        \item \textit{Departure} consequences correspond to the departure of team members from the project or organization.
        This may disrupt continuity, knowledge transfer, and team dynamics.
        \item \textit{Conflicts} consequences correspond to the clashes among individuals or groups.
        This may obstruct collaboration, communication, and progress.
        \item \textit{Frustration and low morale} consequences correspond to the state of reduced motivation, satisfaction, or engagement among team members.
        This may result in decreased productivity and increased staff turnover.
    \end{itemize}
    \item \textbf{Company effects} consequences impact the company as a whole.
    \begin{itemize}
        \item \textit{Damaged reputation} consequences correspond to the negative impact on the organization's image and trustworthiness.
        This may happen due to publicized failures and scandals.
        \item \textit{Toxic environment} consequences correspond to the work culture characterized by hostility, lack of support, excessive pressure, or unhealthy dynamics.
        This may affect employee well-being, teamwork, and overall productivity.
    \end{itemize}
    \item Other forms of impact (please specify)
\end{itemize}

At the end of each section, we provide participants an open comment box to share any additional information or insights regarding the problem. 
The average time required for completing the survey is approximately 15 minutes.

\subsection{Processing and analysis}
After closing the survey for public, we downloaded all recorded responses, including timestamps indicating the survey start and completion times.
These records include both fully completed surveys and partially filled ones.
We filter out the latter during data processing. 
The first step involves filtering out records from unsuitable candidates who started the survey but are not considered eligible. 
Following the criteria outlined in the previous section, we exclude these candidates, leaving \nsurveyOne records available for analysis.
In the subsequent processing stage, we focus on aggregating information provided by participants regarding their experience with \ups. 
Additionally, we match comments provided in the "other" open text boxes with existing response options in the frequency and consequences questions. 
In cases where we find no match, we preserve the comments as originally entered.

The aim of this survey is to gather supplementary data for statistical analysis of \ups.
Firstly, we calculate the percentage of reported frequencies for each pattern. 
It should be noted that some participants indicate that they have never encountered certain patterns or consider them beyond their domain of responsibility. 
Consequently, the number of votes for frequency-related questions differs from the number of votes for significance and consequences questions. 
Secondly, we calculate the percentage of reported significances for each \up. 
Lastly, we determine the percentage of reported consequences for each pattern.
Subsequently, we highlight consequences reported by over 75\% of respondents.
Finally, we compute ratios for respondents who encountered a specific number of patterns out of the five presented to them (ranging from 0 to 5). 
Additionally, we conduct these statistical calculations for distinct groups, including individuals in managerial roles, respondents with varying levels of work experience, and those who had led more than three projects.
The results and conclusions are presented in~\Cref{sec:results_problems}.

\observation{
Ultimately, we extend the list of identified \ups with the following information:
\begin{itemize}
    \item Average frequency;
    \item Average significance;
    \item Average occurrence rate of each consequence.
    The consequence list was compiled during the interviews.
\end{itemize}
}
\section{Results and Analysis from the \surveyOne Survey}\label{sec:results_problems}
After conducting \ninters interviews, we gathered a comprehensive list of \nprobs \ups. 
To obtain quantitative data, we conduct the \surveyOne survey, seeking responses from a larger pool of participants. 
The survey yielded an average of 51 responses for each pattern, ranging from a minimum of 39 to a maximum of 64.
Unfortunately, we cannot present detailed information and statistics for all \ups we study and stay within the reasonable paper page limit.
Therefore, in~\Cref{sec:results_problems_stats}, we present the gathered information and statistics for the ten \ups that we select for the tool research in the \surveyTwo survey. 
These ten patterns, in our opinion from an industrial perspective, have the most realistic and innovative tool and feature proposals from the interviewees.
For a comprehensive overview of all \nprobs \ups, please refer to Appendix A\footnote{Online Appendix: \url{https://github.com/KatyaKos/Undesirable-Patterns/blob/main/Undesirable_Patterns_Appendix.pdf}}.
In~\Cref{sec:results_problems_analysis}, we analyze the survey encompassing all \nprobs patterns and summarize the IT practitioners' perceptions and pain points of various \ups. 

\subsection{Results of the survey}\label{sec:results_problems_stats}
\begin{figure}[htp]
\centering
\includegraphics[width=0.98\columnwidth]{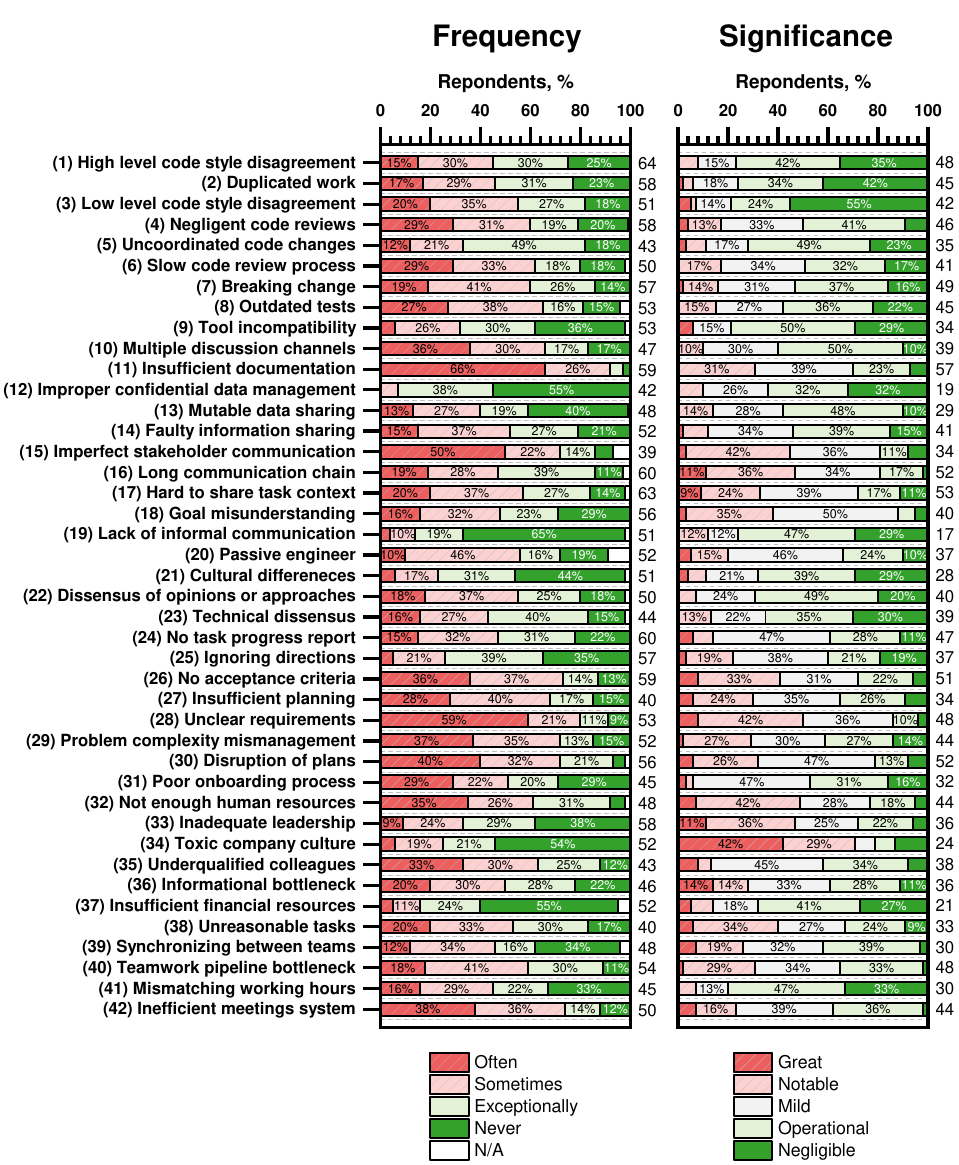}
\caption{Frequency and significance of \ups reported in the \surveyOne survey. Numbers next to the coloured grid show the total number of votes.}
\label{fig:freq_sig_results}
\end{figure}
Tables included in this section present survey results of only ten \ups that, in our opinion from an industrial perspective, have the most realistic and innovative tool and feature proposals from the interviewees.
However, the results that we describe below encompass all \nprobs \ups presented in Appendix A\footnote{Online Appendix: \url{https://github.com/KatyaKos/Undesirable-Patterns/blob/main/Undesirable_Patterns_Appendix.pdf}} \added{and are outlined in} \Cref{fig:freq_sig_results}. 

For each pattern, we collect statistics outlining its properties.
For every frequency, significance, and consequence answer choice offered to participants in the survey, we calculate the percentages of respondents who chose each option.
For the sake of brevity, we are using shorter descriptions in the tables instead of the full problem frequency, significance and consequences descriptions that we present to survey participants.

According to Appendix A.2, \textit{Insufficient documentation} is the most frequent \up, with 66\% of survey participants indicating its frequent occurrence ("Often"), the highest among all patterns.
Moreover, only 3\% report it as never happening, and 5\% label it as exceptionally rare, both being the lowest percentages for these response options among all patterns. 
This observation aligns with the findings from our exploratory interviews, where 32 out of \ninters participants highlight problems related to documentation.
%As presented in Appendix A.4, the most common consequence associated with this pattern is the loss of time on communication.
%This is a logical outcome, since the absence of up-to-date and comprehensive documentation often requires discussions with colleagues responsible for the relevant code or those who have previously worked on it.

\textit{Unclear requirements} emerges as the second most frequent among the \ups, with 59\% of respondents indicating its occurrence as "Often", followed by 11\% for "Exceptionally", and only 9\% for "Never".
Interestingly, this pattern has the highest number of frequently reported consequences (four occurrence scores over 75\%).

The least frequent patterns are \textit{Improper confidential data management}, \textit{Lack of informal communication}, \textit{Insufficient financial resources}, all of which do not have any consequences reported with an occurrence higher than 75\%.

As highlighted in Appendix A.3, \textit{Toxic company culture} stands out as the most significant among the selected \ups.
42\% of respondents indicate its significance as "Great" which is almost 30\% more than the second highest ratio.
As seen in Appendix A.6, this pattern often leads to employee frustration, signalling the urgency for companies to address and rectify toxic cultural elements.
This finding emphasizes the importance of proactive measures to cultivate a healthy and supportive work environment, thereby fostering employee well-being and overall organizational success.

Another curious finding from the survey is that \textit{Low level code style disagreement} is the least significant \up, with 55\% of survey participants indicating its negligible effect on their work, the highest among all patterns.
Additionally, only 2\% of respondents consider its significance as notable, while 5\% label it as great. 
Notably, this pattern lacks reported consequences with an occurrence score exceeding 75\%. 
This is an interesting result, given the multitude of tools, plugins, and rule books dedicated to addressing this pattern.

Taking a broader perspective, let us consider the top ten patterns with the highest scores in "Often" and the top ten with the lowest scores in "Never".
In both cases, seven out of ten patterns fall into the organizational category (by origins). 
Now, shifting the focus to significance scores, let us consider the ten patterns with the highest scores in "Great" and ten with the lowest scores in "Negligible". 
We observe a trend: eight and seven out of ten patterns are organizational. 
\observation{
We see that organizational \ups are both most frequent and most significant of all.
Moreover, organizational patterns have the highest number of consequences with occurrence score exceeding 75\%.
This observation is important, particularly when we note the limited availability of tools designed to identify and address organizational patterns as compared to e.g. tools tailored for technical patterns.
}

\newpage
\noindent
Frequency codes (\Cref{tab:results_problems_stats_freq}):\\
\textit{Often}: I often encounter this problem\\
\textit{Sometimes}: I encounter this problem from time to time, but not very often\\
\textit{Exceptionally}: I've encountered this problem, but it was an unusual situation\\
\textit{Never}: I've never encountered this problem\\
\textit{N/A}: This problem is outside my domain of responsibility\\

\noindent
\begin{table}
\small
\begin{tabular}{ccccccc}
\toprule
\textbf{Pattern} & \multicolumn{5}{c}{\textbf{Frequency (\%)}} & \textbf{Total} \\\cmidrule(lr){2-6}
\textbf{name} & \textbf{Often} & \textbf{Sometimes} & \textbf{Exceptionally} & \textbf{Never} & \textbf{N/A} & \textbf{votes} \\\midrule
\rowcolor{gray} \begin{tabular}{@{}c@{}}(4) Negligent\\code reviews\end{tabular} & 29 & 31 & 19 & 20 & 1 & 58\\
\begin{tabular}{@{}c@{}}(5) Uncoordinated\\code changes\end{tabular} & 12 & 21 & \underline{\textbf{49}} & 18 & 0 & 43\\
\rowcolor{gray} \begin{tabular}{@{}c@{}}(6) Slow code\\review process\end{tabular} & 29 & 33 & 18 & 18 & 2 & 50\\
\begin{tabular}{@{}c@{}}(8) Outdated\\tests\end{tabular} & 27 & 38 & 16 & 15 & 4 & 53\\
\rowcolor{gray} \begin{tabular}{@{}c@{}}(10) Multiple\\discussion\\channels\end{tabular} & 36 & 30 & 17 & 17 & 0 & 47\\
\begin{tabular}{@{}c@{}}(11) Insufficient\\documentation\end{tabular} & \underline{\textbf{66}} & 26 & 5 & 3 & 0 & 59\\
\rowcolor{gray} \begin{tabular}{@{}c@{}}(24) No task\\progress report\end{tabular} & 15 & 32 & 31 & 22 & 0 & 60\\
\begin{tabular}{@{}c@{}}(28) Unclear\\requirements\end{tabular} & 59 & 21 & 11 & 9 & 0 & 53\\
\rowcolor{gray} \begin{tabular}{@{}c@{}}(31) Poor\\onboarding\\process\end{tabular} & 29 & 22 & 20 & \underline{\textbf{29}} & 0 & 45\\
\begin{tabular}{@{}c@{}}(42) Inefficient\\meetings system\end{tabular} & 38 & \underline{\textbf{36}} & 14 & 12 & 0 & 50\\\bottomrule
\end{tabular}
\caption{\label{tab:results_problems_stats_freq}Reported frequency of 10 patterns chosen for the tool ideas evaluation. The highest scores for each frequency are highlighted.}
\end{table}

\newpage
\noindent
Significance codes (\Cref{tab:results_problems_stats_sig}):\\
\textit{Great}: This problem greatly impacts work-related processes (the project is closed or postponed indefinitely, some of the team members leave the team due to conflicts or general dissatisfaction)\\
\textit{Notable}: This problem significantly impacts work-related processes (release deadlines are moved significantly as compared to the release cycle, some critical features are not shipped in time)\\
\textit{Mild}: This problem mildly impacts work-related processes (release deadlines are slightly moved, some features are postponed to be released later, managers participate in resolving the problem)\\
\textit{Operational}: This problem affects the workflow, but does not impact work-related processes (release deadlines are not moved, no overtime work is required, no need to involve managers or other outside people to address the issue). Intermediate goals, however, may have to be changed\\
\textit{Negligible}: This problem hardly affects the workflow or does not affect it at all

\noindent
\begin{table}
\small
\centering
% \resizebox{5in}{!}{%
\begin{tabular}{ccccccc}

\toprule
\textbf{Pattern} & \multicolumn{5}{c}{\textbf{Significance (\%)}} & \textbf{Total} \\\cmidrule(lr){2-6}
\textbf{name} & \textbf{Great} & \textbf{Notable} & \textbf{Mild} & \textbf{Operational} & \textbf{Negligible} & \textbf{votes} \\\midrule
\rowcolor{gray} \begin{tabular}{@{}c@{}}(4) Negligent\\code reviews\end{tabular} & 4 & 13 & 33 & 41 & 9 & 46\\
\begin{tabular}{@{}c@{}}(5) Uncoordinated\\code changes\end{tabular} & 3 & 8 & 17 & \underline{\textbf{49}} & \underline{\textbf{23}} & 35\\
\rowcolor{gray} \begin{tabular}{@{}c@{}}(6) Slow code\\review process\end{tabular} & 0 & 17 & 34 & 32 & 17 & 41\\
\begin{tabular}{@{}c@{}}(8) Outdated\\tests\end{tabular} & 0 & 15 & 27 & 36 & 22 & 45\\
\rowcolor{gray} \begin{tabular}{@{}c@{}}(10) Multiple\\discussion\\channels\end{tabular} & 0 & 10 & 30 & 50 & 10 & 39\\
\begin{tabular}{@{}c@{}}(11) Insufficient\\documentation\end{tabular} & 0 & 31 & 39 & 23 & 7 & 57\\
\rowcolor{gray} \begin{tabular}{@{}c@{}}(24) No task\\progress report\end{tabular} & 6 & 8 & \underline{\textbf{47}} & 28 & 11 & 47\\
\begin{tabular}{@{}c@{}}(28) Unclear\\requirements\end{tabular} & \underline{\textbf{8}} & \underline{\textbf{42}} & 36 & 10 & 4 & 48\\
\rowcolor{gray} \begin{tabular}{@{}c@{}}(31) Poor\\onboarding\\process\end{tabular} & 3 & 3 & \underline{\textbf{47}} & 31 & 16 & 32\\
\thead{(42) Inefficient\\meetings system} & 7 & 16 & 39 & 36 & 2 & 44\\\bottomrule
\end{tabular}
% }
\caption{\label{tab:results_problems_stats_sig}Reported significance of 10 patterns chosen for the tool ideas evaluation. The highest scores for each significance are highlighted.}
\end{table}

\newpage
\begin{table}[!p]
\small
\caption{\label{tab:results_problems_stats_cons}Reported consequences of 10 patterns chosen for the tool ideas evaluation. Scores over 75\% are highlighted.}
\begin{tabular}{ccccc}

\toprule
\textbf{Pattern} & \multicolumn{3}{c}{\textbf{Wasting time consequences (\%)}} & \textbf{Total}\\\cmidrule(lr){2-4}
\textbf{name} & \textbf{Waiting} & \textbf{Fixing or redoing} & \textbf{Communication} & \textbf{votes} \\\midrule
\rowcolor{gray} \begin{tabular}{@{}c@{}}(4) Negligent\\code reviews\end{tabular} & 56 & \underline{\textbf{82}} & 51 & 46 \\
\begin{tabular}{@{}c@{}}(5) Uncoordinated\\code changes\end{tabular} & 35 & 74 & 52 & 35 \\
\rowcolor{gray} \begin{tabular}{@{}c@{}}(6) Slow code\\review process\end{tabular} & \underline{\textbf{93}} & 33 & 48  & 41 \\
\begin{tabular}{@{}c@{}}(8) Outdated tests\end{tabular} & 33 & 74 & 38 & 45 \\
\rowcolor{gray} \begin{tabular}{@{}c@{}}(10) Multiple\\discussion channels\end{tabular} & 49 & 43 & \underline{\textbf{89}} & 39 \\
\begin{tabular}{@{}c@{}}(11) Insufficient\\documentation\end{tabular} & 58 & 55 & \underline{\textbf{84}} & 57 \\
\rowcolor{gray} \begin{tabular}{@{}c@{}}(24) No task\\progress report\end{tabular} & 60 & \underline{\textbf{78}} & 51 & 47 \\
\begin{tabular}{@{}c@{}}(28) Unclear\\requirements\end{tabular} & 66 & \underline{\textbf{81}} & \underline{\textbf{79}} & 48\\
\rowcolor{gray} \begin{tabular}{@{}c@{}}(31) Poor\\onboarding process\end{tabular} & 39 & 57 & \underline{\textbf{82}} & 32 \\
\begin{tabular}{@{}c@{}}(42) Inefficient\\meetings system\end{tabular} & 42 & 28 & \underline{\textbf{79}} & 44 \\\bottomrule
\end{tabular}

% \vspace{10pt}

\begin{tabular}{ccccc}
\toprule
\textbf{Pattern} & \multicolumn{4}{c}{\textbf{Project development impact consequences (\%)}} \\\cmidrule(lr){2-5}
\textbf{name} & \textbf{Cancelled} & \textbf{Insupportable} & \textbf{Delayed} & \textbf{Suboptimal} \\\midrule
\rowcolor{gray} \begin{tabular}{@{}c@{}}(4) Negligent\\code reviews\end{tabular} & 2 & 22 & 36 & 67 \\
\begin{tabular}{@{}c@{}}(5) Uncoordinated\\code changes\end{tabular} & 3 & 0 & 42 & 26 \\
\rowcolor{gray} \begin{tabular}{@{}c@{}}(6) Slow code\\review process\end{tabular} & 0 & 3 & 28 & 28 \\
\begin{tabular}{@{}c@{}}(8) Outdated tests\end{tabular} & 0 & 15 & 59 & 51 \\
\rowcolor{gray} \begin{tabular}{@{}c@{}}(10) Multiple\\discussion channels\end{tabular} & 0 & 8 & 27 & 24 \\
\begin{tabular}{@{}c@{}}(11) Insufficient\\documentation\end{tabular} & 4 & 16 & 53 & 44 \\
\rowcolor{gray} \begin{tabular}{@{}c@{}}(24) No task\\progress report\end{tabular} & 7 & 9 & 51 & 31 \\
\begin{tabular}{@{}c@{}}(28) Unclear\\requirements\end{tabular} & 9 & 9 & \underline{\textbf{75}} & 60 \\
\rowcolor{gray} \begin{tabular}{@{}c@{}}(31) Poor\\onboarding process\end{tabular} & 0 & 11 & 21 & 29 \\
\begin{tabular}{@{}c@{}}(42) Inefficient\\meetings system\end{tabular} & 2 & 9 & 47 & 37 \\\bottomrule
\end{tabular}
\end{table}

\begin{table}[!p]
\begin{tabular}{ccccc}
\toprule
\textbf{Pattern} & \multicolumn{3}{c}{\textbf{Team effects consequences (\%)}} & \textbf{Total}\\\cmidrule(lr){2-4}
\textbf{name} & \textbf{Departure} & \textbf{Conflicts} & \textbf{Frustration} & \textbf{votes} \\\midrule
\rowcolor{gray} \begin{tabular}{@{}c@{}}(4) Negligent\\code reviews\end{tabular} & 9 & 37 & 42 & 46 \\
\begin{tabular}{@{}c@{}}(5) Uncoordinated\\code changes\end{tabular} & 6 & 23 & 26 & 35 \\
\rowcolor{gray} \begin{tabular}{@{}c@{}}(6) Slow code\\review process\end{tabular} & 3 & 18 & 48  & 41 \\
\begin{tabular}{@{}c@{}}(8) Outdated tests\end{tabular} & 13 & 21 & 33 & 45 \\
\rowcolor{gray} \begin{tabular}{@{}c@{}}(10) Multiple\\discussion channels\end{tabular} & 0 & 41 & 51 & 39 \\
\begin{tabular}{@{}c@{}}(11) Insufficient\\documentation\end{tabular} & 9 & 11 & 56 & 57 \\
\rowcolor{gray} \begin{tabular}{@{}c@{}}(24) No task\\progress report\end{tabular} & 7 & 47 & 47 & 47 \\
\begin{tabular}{@{}c@{}}(28) Unclear\\requirements\end{tabular} & 19 & 36 & \underline{\textbf{77}} & 48\\
\rowcolor{gray} \begin{tabular}{@{}c@{}}(31) Poor\\onboarding process\end{tabular} & 7 & 11 & 61 & 32 \\
\begin{tabular}{@{}c@{}}(42) Inefficient\\meetings system\end{tabular} & 21 & 21 & 70 & 44 \\\bottomrule
\end{tabular}

% \vspace{10pt}

\begin{tabular}{ccc}
\toprule
\textbf{Pattern} & \multicolumn{2}{c}{\textbf{Company effects consequences (\%)}} \\\cmidrule(lr){2-3}
\textbf{name} & \textbf{Damaged reputation} & \textbf{Toxic environment} \\\midrule
\rowcolor{gray} \begin{tabular}{@{}c@{}}(4) Negligent\\code reviews\end{tabular} & 22 & 11 \\
\begin{tabular}{@{}c@{}}(5) Uncoordinated\\code changes\end{tabular} & 3 & 16 \\
\rowcolor{gray} \begin{tabular}{@{}c@{}}(6) Slow code\\review process\end{tabular} & 10 & 8 \\
\begin{tabular}{@{}c@{}}(8) Outdated tests\end{tabular} & 21 & 21 \\
\rowcolor{gray} \begin{tabular}{@{}c@{}}(10) Multiple\\discussion channels\end{tabular} & 8 & 16 \\
\begin{tabular}{@{}c@{}}(11) Insufficient\\documentation\end{tabular} & 13 & 13 \\
\rowcolor{gray} \begin{tabular}{@{}c@{}}(24) No task\\progress report\end{tabular} & 5 & 20 \\
\begin{tabular}{@{}c@{}}(28) Unclear\\requirements\end{tabular} & 26 & 26 \\
\rowcolor{gray} \begin{tabular}{@{}c@{}}(31) Poor\\onboarding process\end{tabular} & 7 & 7 \\
\begin{tabular}{@{}c@{}}(42) Inefficient\\meetings system\end{tabular} & 12 & 16 \\\bottomrule
\end{tabular}
\end{table}

\FloatBarrier
\subsection{Analysis}\label{sec:results_problems_analysis}

\noindent
\begin{table}[!htpb]
\small
\begin{tabular}{cccccccc}
\toprule
 & \multicolumn{6}{c}{\thead{\textbf{Number of problems met}\\\textbf{(out of 5, \%)}}} & \thead{\textbf{Total}\\\textbf{participants}}\\\cmidrule(lr){2-7}
 & \textbf{0} & \textbf{1} & \textbf{2} & \textbf{3} & \textbf{4} & \textbf{5} &  \\\midrule
\rowcolor{gray} \textbf{All}  & 2 & 4 & 8 & 19 & 29 & 38 & \nsurveyOne\\
\textbf{Management} & 1 & 3 & 8 & 16 & 32 & 40  & 122\\
\rowcolor{gray} \textbf{$<$4 years of experience} & 3 & 3 & 19 & 26 & 32 & 17 & 31 \\
\textbf{7+ years of experience} & 2 & 4 & 7 & 17 & 29 & \underline{\textbf{41}} & 312 \\
\rowcolor{gray} \textbf{Led 3+ projects} & 1 & 6 & 12 & 23 & 27 & 31 & 214\\\bottomrule
\end{tabular}
\caption{\label{tab:results_problems_analysis}Percentages of people who met a specific number of \ups out of five presented to them (chose any frequency option except for "Never" and "N/A").}
\end{table}

All \nsurveyOne survey respondents evaluate five \ups that are randomly assigned to them to ensure random and uniform distribution of all the patterns.
These five \ups come from the full list of 42 \ups we identify in the previous section.
During the evaluation process, we ask the respondents to specify the frequency at which they encountered each \up.
If any answer option except for "Never" and "N/A" is selected, it indicates that the participants had experienced the pattern at least once during their professional journey.
In ~\Cref{tab:results_problems_analysis}, we calculate the ratio of participants who had come across a specific number of patterns, ranging from 0 to 5. 
We further extend this analysis to particular participant categories: (1) individuals in managerial roles such as CIOs, CEOs, and CTOs, Team Leads, or Product Managers, (2) participants with less than four years of experience, (3) participants with a minimum of seven years of experience, and (4) those who have led at least three team projects.

\observation{
Except for one group, those with less than four years of experience, in every other group the majority of respondents encountered all five \ups.
However, even among individuals with less than four years of experience, the majority of respondents met three or four patterns.
}
Predictably, the group of more experienced participants shows the highest percentage of individuals (41\%) who have encountered all five patterns.
On average, participants have come across 3.8 out of the five \ups in their professional experiences.
Across all groups, only 1-2\% of respondents have never encountered any of the suggested patterns, translating to a range of 1 to 6 individuals.
\section{\surveyTwo\ survey design}\label{sec:design_tools}
\begin{figure}[htp]
\centering
\includegraphics[width=0.98\columnwidth]{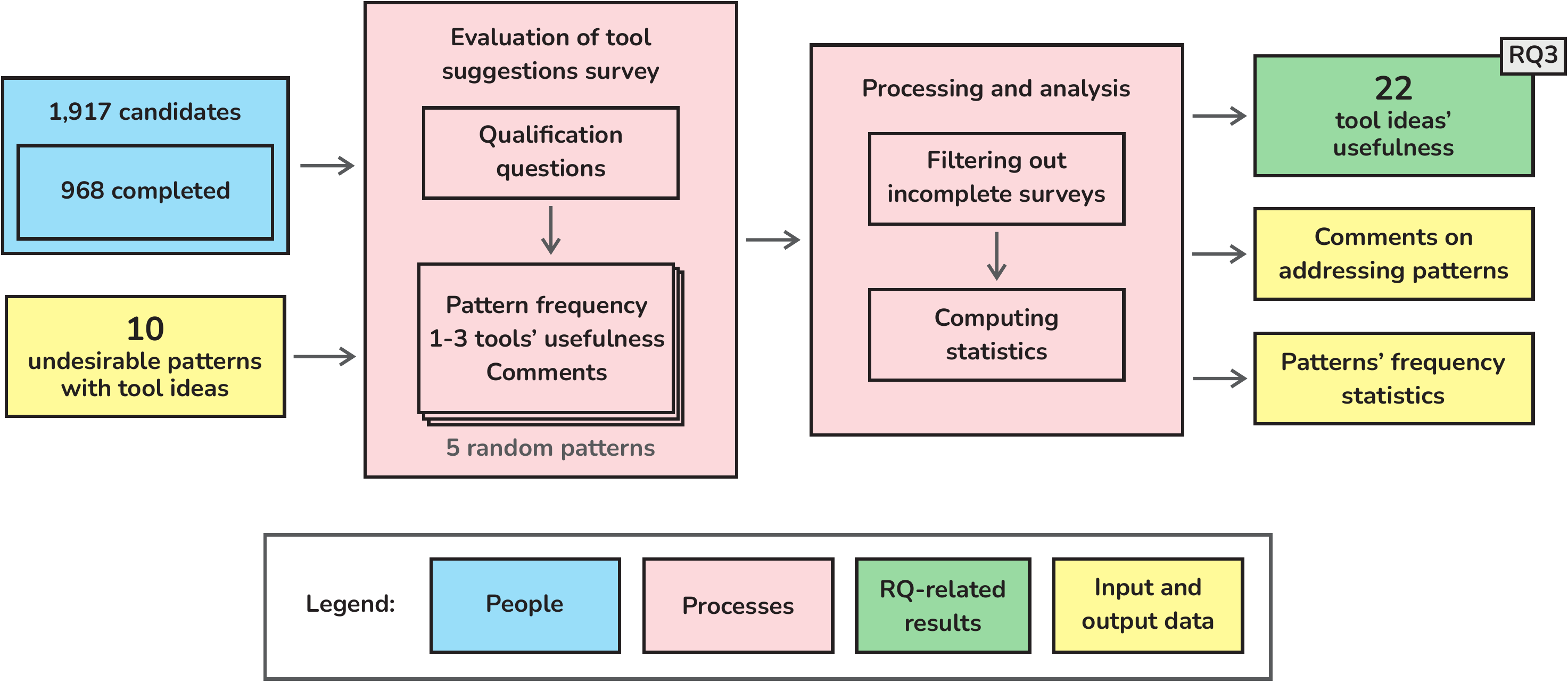}
\caption{Pipeline of the \surveyTwo survey phase.}
\label{fig:scheme_surv2}
\end{figure}

During the interviews, we gather a list of tool and feature ideas for some of the identified patterns.
\added{We filter out ideas that are too general or nonsensical, e.g. "AI tool that fixes it" or "a feature that blocks your pass if there are any pending pull requests".
However, the remaining list is still too long, and} conducting evaluations for all these ideas with statistical significance would require a substantial number of participants.
In this research, we focus on presenting and evaluating tool ideas for 10 \ups based on the perceived practicality and innovative nature of the tools proposed by the interviewees, considering an industrial perspective.
The main goal of the second survey, \surveyTwo, is to assess the potential usefulness of these tool ideas, providing valuable insights into which tools are worth considering for implementation.
\Cref{fig:scheme_surv2} depicts the pipeline of this phase of the study.
The survey script is accessible in the Appendix  D\footnote{Online Appendix: \url{https://github.com/KatyaKos/Undesirable-Patterns/blob/main/Undesirable_Patterns_Appendix.pdf}}.

Additionally, we inquire about the level of influence each respondent holds within their respective teams when it comes to decision-making regarding new tools and practices.
We collected \nsurveyTwo responses, approximately 400 for each tool suggestion.
More than 600 responses come from individuals who either possess sole authority in this domain within their teams or share this authority with only a small group. 
These respondents hold paramount importance to us in shaping our understanding and decision-making process regarding tool adoption.

\subsection{Recruitment and demographics}
\begin{figure}[htp]
\centering
\subfloat[Specialization (multiple choice question)]{
\includegraphics[width=0.7\columnwidth]{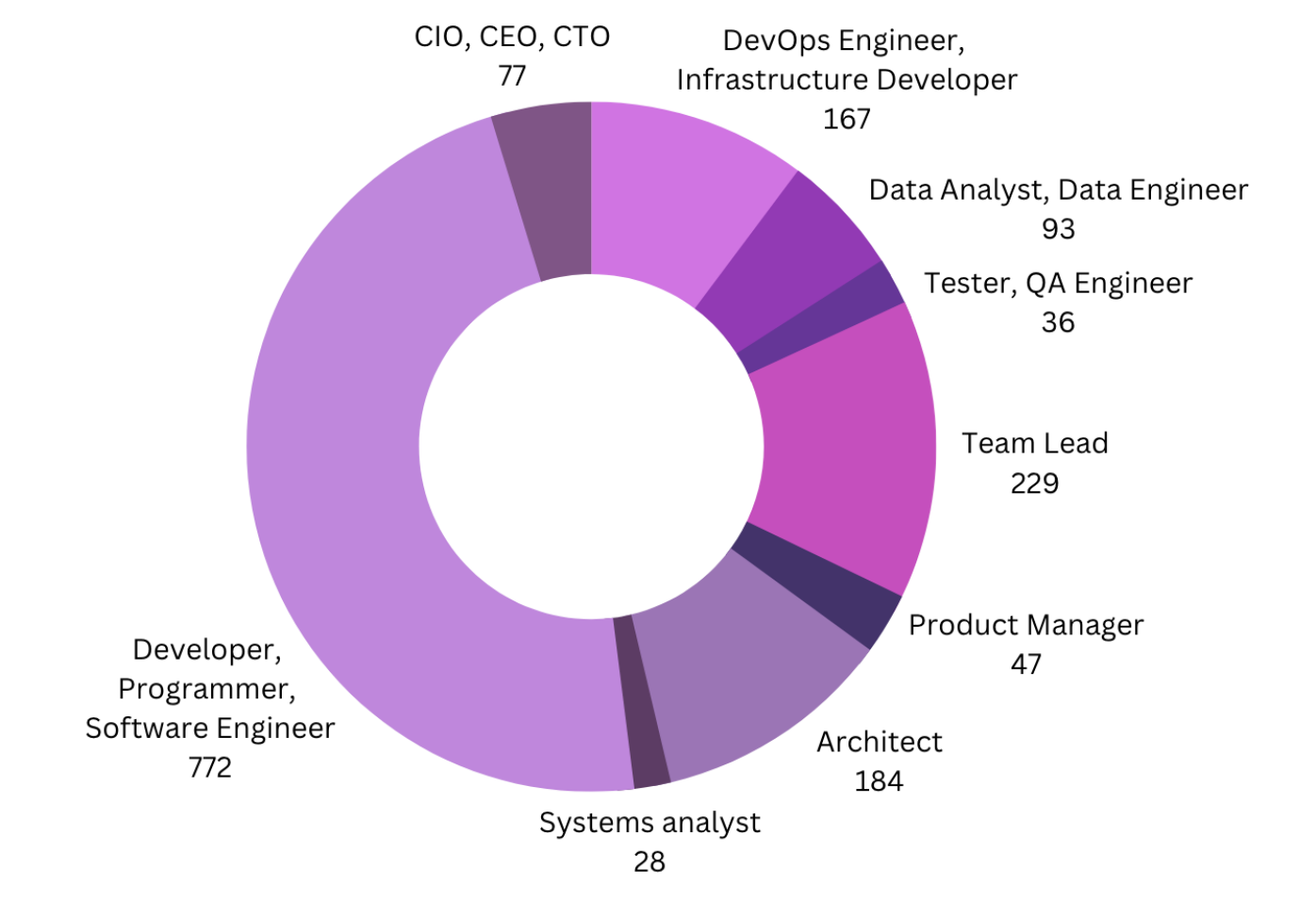}
\label{fig:surv2_dem_role}}
\\
\subfloat[Employment status]{
\includegraphics[width=0.4\columnwidth]{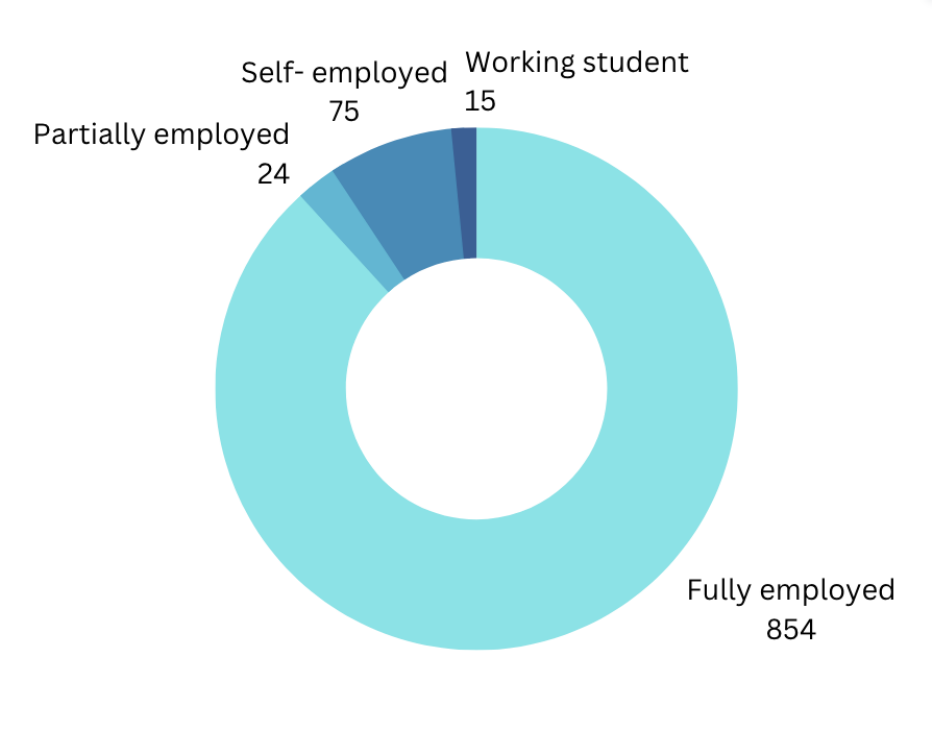}
\label{fig:surv2_dem_empl}}
\subfloat[Years of experience]{
\includegraphics[width=0.4\columnwidth]{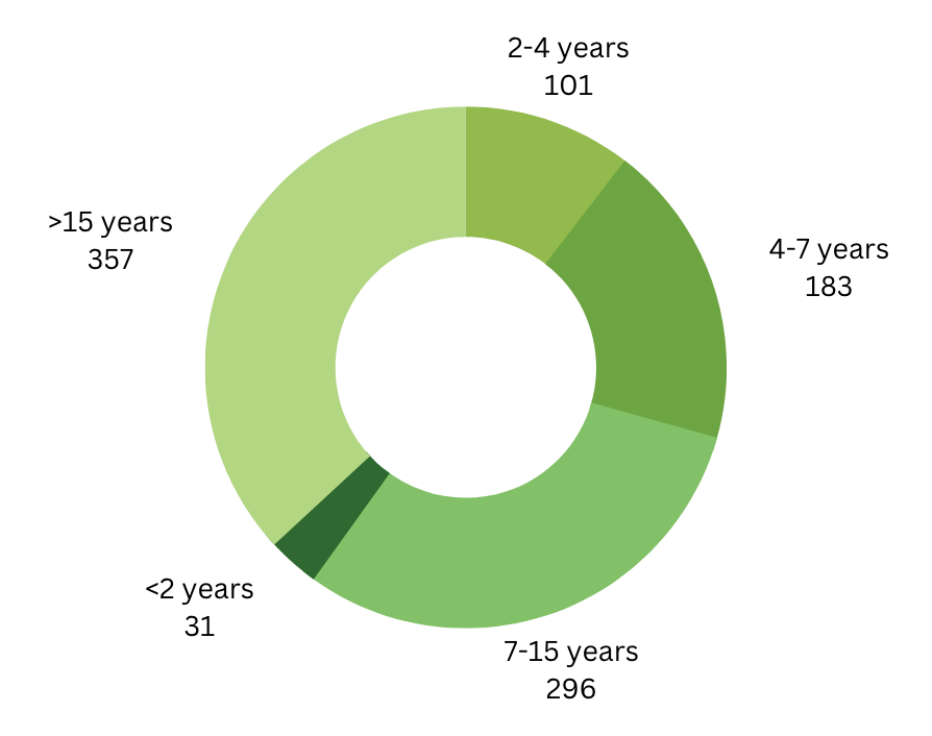}
\label{fig:surv2_dem_exp}}
\\
\subfloat[Participated in team projects]{
\includegraphics[width=0.4\columnwidth]{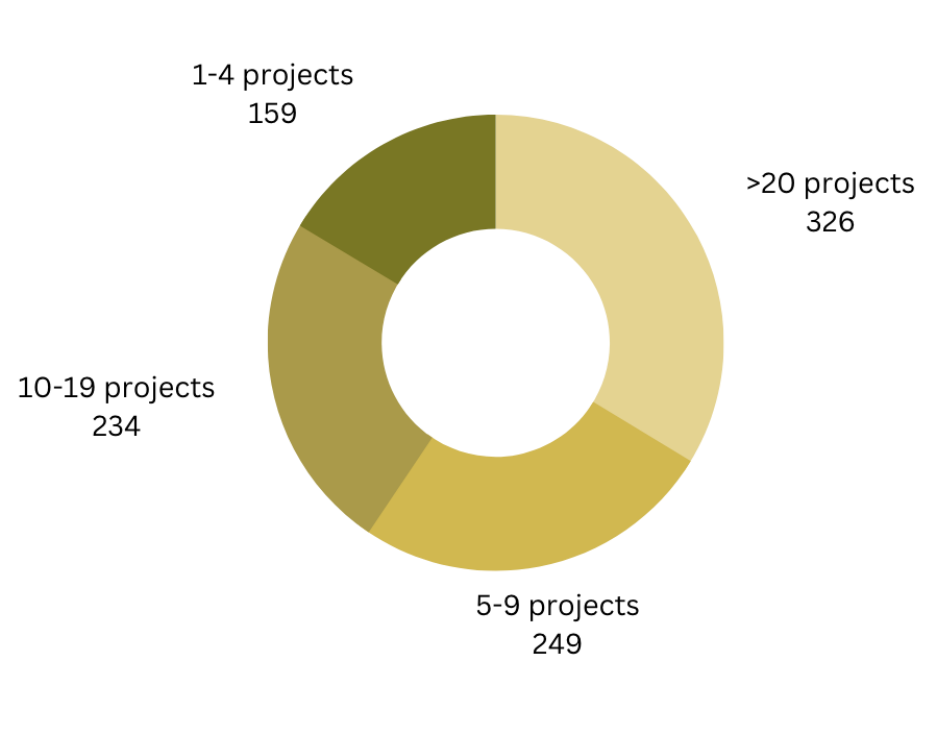}
\label{fig:surv2_dem_projs_part}}
\subfloat[Influence on adopting tools]{
\includegraphics[width=0.4\columnwidth]{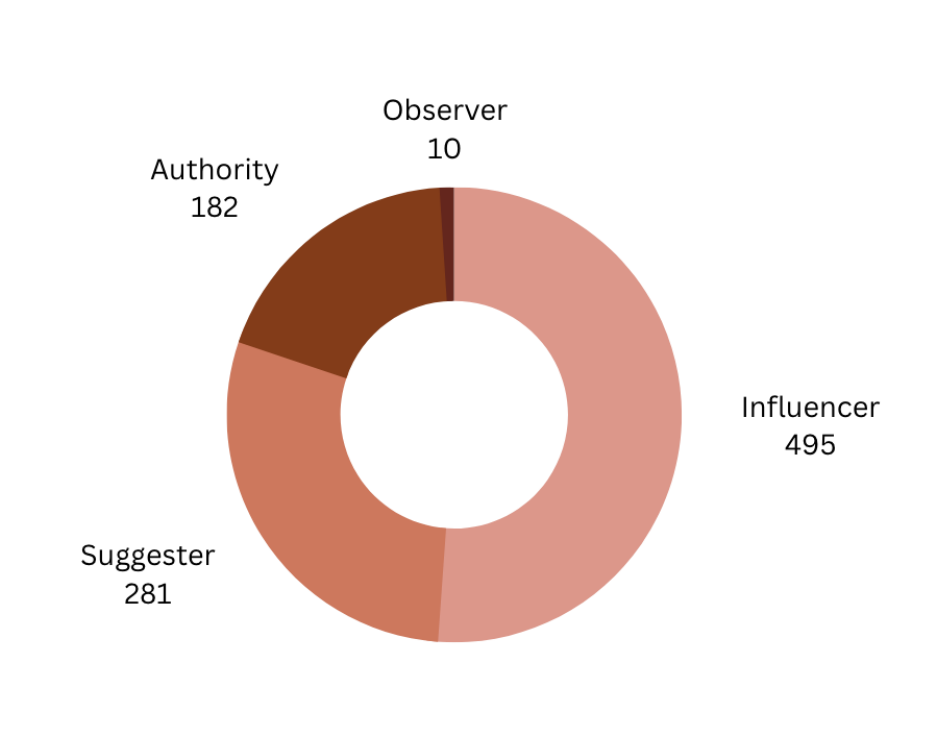}
\label{fig:surv2_dem_vote}}
\caption{Demographic infographics for the \surveyTwo survey with \nsurveyTwo participants.}
\label{fig:surv2_dem}
\end{figure}

The \surveyTwo survey begins with a series of background questions aimed at attracting a diverse pool of participants from various professional backgrounds (\Cref{fig:surv2_dem}). 
The survey is conducted in English.
To ensure the suitability of candidates, the survey employs four filtering questions, similar to those used in the \surveyOne survey.
These questions pertain to employment status (\Cref{fig:surv2_dem_empl}), years of experience (filtering out candidates not employed in IT \Cref{fig:surv2_dem_exp}), current job title (\Cref{fig:surv2_dem_role}), and the number of team projects in which the candidate participated (filtering out candidates with zero projects \Cref{fig:surv2_dem_projs_part}).
Initially, 1917 candidates responded to the first question; however, after undergoing the filtering process, \nsurveyTwo individuals met the specified criteria and successfully completed the main survey.

As part of the survey, candidates are also asked about the level of influence they have regarding the adoption of new tools in their teams (\Cref{fig:surv2_dem_vote}). 
Although this question is not used for filtering, the answers provided valuable insights for later survey analysis.
Particularly, the responses from candidates with higher influence are of special interest, as they can help us predict the potential widespread adoption of specific tools or features.
For the sake of brevity, in the paper we are using following descriptions of influence groups:\\

\noindent
\textit{Authority}: I can decide that our team will use this tool from now on, and it is unlikely to be overruled by someone else\\
\textit{Influencer}: I cannot decide that our team will adopt this tool, but I can strongly influence a discussion about whether or not this tool should be used (e.g. I’m one of 3 people who will make this decision together)\\
\textit{Suggester}: I can suggest using this tool, but I have no impact on whether it will be adopted or not\\
\textit{Observer}: I have no say in these matters\\

We inform participants that the research findings would be shared with the community while ensuring the confidentiality of the information they provided. 
Furthermore, the participants who successfully complete the survey are automatically entered into a lottery with a prize of their choice: a one year \jb All Products Pack subscription or a \$100 Amazon gift card.

\subsection{Survey outline and study protocol}
We conduct the \surveyTwo survey using the Alchemer platform~\cite{alchemer}. 
After the filtering background questions, the survey focuses on specific \ups and the proposed tool ideas to address them. 
Similar to the \surveyOne survey, participants receive explanations for terms "technical problem" and "undesirable pattern" along with instructions for the further questions.
Participants are then presented with five sections of questions, each dedicated to a randomly assigned pattern from the list of 10 patterns we selected for this survey.
This ensures the random and uniform distribution of the patterns and their related tools among all participants. 
On average, we received 418 responses for each tool, with a maximum of 476 and a minimum of 355.

In the following, we present the structure of each pattern section, where we maintain a consistent format, with the only difference being the pattern's name and description that marks the beginning of the section.
We then ask the participants about how frequently they encounter the particular pattern in their professional experience, using the same answer options as in the \surveyOne survey.
Participants who indicate that the pattern is outside their domain of responsibility or that they have never encountered it do not proceed to evaluate the corresponding tools and are directed to the next section.
After providing their frequency assessment for the pattern, participants are presented with one to three tool ideas that can potentially help address the identified pattern.
We present each idea together with its description; we do not name the ideas to avoid the "first impression" bias that often comes with the names.
For each idea, we ask participants to evaluate its usefulness using the following answer options:
\begin{itemize}
    \item Not beneficial at all
    \item Helps to manage the problem a little
    \item Solves a part of the problem
    \item Solves the problem completely
\end{itemize}
Once participants evaluate all the ideas, they are provided with three open-answer text boxes to respond to the following non-obligatory questions:
\begin{enumerate}
    \item If you have another idea for a tool that could help manage the problem mentioned above, please describe it here. 
    \item Please add any comments you may have on the tool concepts presented above (e.g. possible modifications to the tool, or advantages and disadvantages of using it).
    \item Is there anything else you would like to share regarding the problem and any possible solutions?
\end{enumerate}
The average time required for completing the survey is approximately 12 minutes.

\subsection{Processing and analysis}
The data processing for the \surveyTwo survey follows a similar approach to that of the \surveyOne survey. 
We start with downloading responses, analyzing timestamps, and filtering out unsuitable candidates.
Finally, we match comments from the "other" text box in the pattern frequency questions with existing answer options.
The primary objective of this survey is to conduct initial research on the usefulness of the tool ideas collected during the interview process.
This evaluation helps us to identify which ideas are worth presenting to product teams which develop tools for software engineers.
For each idea, we gather all the ratings provided by survey participants and calculate their arithmetic mean for different groups of practitioners based on their influence in adopting new tools. 
We consider opinions from individuals with greater influence to be more valuable from an industrial perspective.

Additionally, we revisit the pattern frequencies, taking into account the larger number of participants in the \surveyTwo survey compared to the \surveyOne survey: an average of 51 responses per pattern versus 418 responses.
This allows for more in-depth statistical analysis and the identification of meaningful correlations. 
We evaluate the probabilities of participants encountering a certain number of patterns out of the five given, and further examine the results based on voters' roles, experience, and number of projects. 
We present all the findings in~\Cref{sec:results_tools}.

\observation{
We design the \surveyTwo survey to evaluate the potential usefulness of 22 tools that were proposed to address the 10 selected \ups, and to collect additional ideas and comments regarding ways to address these issues. 
Additionally, the survey allows us to gather more data on the frequency of these 10 patterns.
}
\section{Results and Analysis from \surveyTwo Survey}\label{sec:results_tools}
After conducting \ninters exploratory interviews, we compile a list of \ups and 22 associated tool ideas that could potentially address these issues. 
Our intention in this survey is to assess these tool ideas to determine which ones would receive the most interest for potential implementation. 
However, given the considerable number of ideas, acquiring statistically significant feedback for all of them poses a challenge. 
We decided to select 10 \ups, along with what we perceive to be the most viable and innovative tool and feature concepts from an industrial perspective for each pattern.
Furthermore, the insights gained from the \surveyOne survey allow us to identify patterns that exhibit higher frequencies and significance. 
This information proved instrumental in resolving disagreements between the first two authors concerning the selection of the ten patterns and their associated tools for evaluation in the subsequent \surveyTwo survey.
The survey yielded an average of 416 responses for each tool, ranging from a maximum of 476 to a minimum of 354.

\subsection{Tools Ideas Evaluation}\label{sec:results_tools_ideas}
Here, we present a list of tool ideas chosen for evaluation in the \surveyTwo survey. 
We assigned a code to each idea which we later use in \Cref{tab:results_tools_ideas_eval}.
\\~\\
\noindent
\textbf{(4) Negligent code reviews (NCR)}
\begin{itemize}
    \item{\textbf{NCR-1}} A tool that tracks how much time has been spent on code review and alerts all stakeholders if it’s insufficient.
    \item{\textbf{NCR-2}} A tool that forces reviewers to write at least X words or spend at least Y minutes on the review.
\end{itemize}
\textbf{(5) Uncoordinated code changes (UCC)}
\begin{itemize}
    \item{\textbf{UCC-1}} A tool that tracks the commit times of various developers and analyzes the frequency of commits and merges affecting a particular part of the codebase. 
The tool warns engineers if they try to commit to a frequently changed part of the codebase regardless of the particular branch. 
This notifies them to proceed with extra caution.
    \item{\textbf{UCC-2}} A tool that allows users to favorite some specific files or methods inside their Git-hosting platform, and alerts them if those files or methods were changed by someone else.
    \item{\textbf{UCC-3}} An IDE tool that allows for a superficial temporary merge of another branch into the current one, and views the code of both branches together as if in a merge conflict resolver.
\end{itemize}
\textbf{(6) Slow code review process (SCRP)}
\begin{itemize}
    \item{\textbf{SCRP-1}} A feature that allows review requesters to set a timer that sends a notification to assignees if the review has not been completed in time. 
    \item{\textbf{SCRP-2}} A feature that provides an extension to the usual merge process. 
First, when the reviewer and the committer agree on all but a few minor details, such as function naming, the code is “pre-merged”. 
This means it has been merged into the codebase with some kind of warning. 
Then, after all of the issues are resolved, the warning is removed and the review is closed. 
This would allow the team to add working and tested features to the product faster.
    \item{\textbf{SCRP-3}} An IDE extension that allows for code reviews in an IDE. 
The reviewer checks out someone else's branch and opens the pending code review on that branch. 
The reviewer then makes notes and comments inside the IDE.
\end{itemize}
\textbf{(8) Outdated tests (OT)}
\begin{itemize}
    \item{\textbf{OT-1}} An IDE feature that links the tests to the code and alerts the committer if the code was altered, but the tests were not updated, like when a method functionality was extended but the test was not updated to check the new functionality.
    \item{\textbf{OT-2}} A tool that stores the testing results and checks whether the test statuses have changed after a new commit. 
It then produces a report that shows which of the tests started failing or working after the commit.
\end{itemize}
\textbf{(10) Multiple discussion channels (MDC)}
\begin{itemize}
    \item{\textbf{MDC-1}} A tool that checks whether several meeting participants are assigned to the same ticket whose description or name matches the meeting’s name. 
If this is true, the tool assumes that this ticket is relevant to the meeting and sends the meeting participants a suggestion to update the ticket if necessary.
    \item{\textbf{MDC-2}} A feature that allows for the linking of chats and threads in a messenger, such as Slack, with specific issues in the issue tracker (e.g. a thread in Slack is linked to an issue on Jira).
\end{itemize}
\textbf{(11) Insufficient documentation (ID)}
\begin{itemize}
    \item{\textbf{ID-1}} 
    A tool that gives the users the ability to cross-link code, issues, chats, and documentation (e.g. a feature in Slack which enables linking of threads to a specific issue on Jira or YouTrack). 
    The tool then extracts the topic from the ticket name and generates documentation by collecting data from all of the cross-links. 
    Engineers can then update the documentation accordingly. 
    \item{\textbf{ID-2}} 
    A tool that sends code authors alerts when they commit a new piece of code and forget to create documentation for it. 
    It also sends notifications to code owners if the corresponding piece of documentation was last updated long before the last code update.
    \item{\textbf{ID-3}} 
    A feature that allows for the creation of documentation right in your IDE. 
    You select a piece of code, press a button, and a new window will appear, in which you can write documentation for this code snippet. 
    This documentation will be uploaded to the team’s storage and automatically linked with the commit.
\end{itemize}
\textbf{(24) No task progress report (NTPR)} 
\begin{itemize}
    \item{\textbf{NTPR-1}} A tool that allows you to create a short description of your current task, so that your colleagues are aware of what you are working on. 
Every day, users flag tasks they are working on, close them once completed, and then create new ones. 
The tool’s simple notepad-like design allows you to quickly update everyone on the status of your tasks, and see how far your colleagues are progressing on your team’s project.
\end{itemize}
\textbf{(28) Unclear requirements (UR)}
\begin{itemize}
    \item{\textbf{UR-1}} A feature that analyzes created tickets.
The ticket should not be empty or too complex (e.g. long compound sentences).
Its name should also match the content.
    \item{\textbf{UR-2}} A tool that structures the writing of tickets.
A particular implementation can be a chatbot which prompts users with specific questions and forms to complete (e.g. “What is the goal?” and “When is the deadline?”). 
It then produces a ticket with all of the submitted information.
\end{itemize}
\textbf{(31) Poor onboarding process (POP)}
\begin{itemize}
    \item{\textbf{POP-1}} A tool to keep all of the answers to frequent onboarding-related issues in one place. 
This would be similar to a separate Stack Overflow~\cite{stackoverflow} workspace for the team or company.
    \item{\textbf{POP-2}} A tool that tracks newcomer commit frequency over the first 6 months of work. 
This data can be used to analyze new engineers’ development pace, and thus deduce the time spent on their onboarding.
\end{itemize}
\textbf{(42) Inefficient meeting system (IMS)}
\begin{itemize}
    \item{\textbf{IMS-1}} A bot that sends notifications X minutes before the meeting ends. 
The exact number is customizable by the meeting creator, and the notifications are similar to those prior to the meeting.
    \item{\textbf{IMS-2}} A tool that analyzes the data for the meeting’s duration. 
It creates diagrams for employees regarding how much time they spend on meetings for each project and notifies the users if they spend a disproportionate amount of time on meetings for certain projects compared to others.
\end{itemize}

\noindent
\begin{table}[!htbp]
\small
\centering
\begin{tabular}{cccccc}
\toprule
& \textbf{Idea} & \multicolumn{4}{c}{\textbf{Idea scores}} \\\cmidrule(lr){3-6}
& \textbf{code} & \textbf{A} & \textbf{AI} & \textbf{AIS} & \textbf{AISO} \\\midrule
\multirow{2}{*}{\begin{tabular}{@{}c@{}}(4) Negligent\\ code reviews\end{tabular}}&  NCR-1 & 0.49 & 0.45 &  0.425 &  0.425 \\
& NCR-2 & 0.395 & 0.335 & 0.33 & 0.33 \\\midrule
\multirow{3}{*}{\begin{tabular}{@{}c@{}}(5) Uncoordinated\\code changes\end{tabular}
}&  UCC-1 &  0.765 & 0.675 &  0.66 &  0.66 \\
& UCC-2 & 0.81 & 0.685 & 0.665 & 0.665 \\
&  UCC-3 & \underline{\textbf{1.03}} & 0.89 & 0.895 &  0.895 \\\midrule
\multirow{3}{*}{\begin{tabular}{@{}c@{}}(6) Slow code\\review process\end{tabular}}& SCRP-1 & 0.765 & 0.745 & 0.73 & 0.74 \\
&  SCRP-2  & 0.62 & 0.61 &  0.625 &  0.625 \\
& SCRP-3 & 0.935 & 0.855 & 0.865 & 0.865 \\\midrule
\multirow{2}{*}{\begin{tabular}{@{}c@{}} (8) Outdated\\tests\end{tabular}}&  OT-1 &  \underline{\textbf{1.01}} &  \underline{\textbf{1.07}} & \underline{\textbf{1.055}} & \underline{\textbf{1.055}} \\
& OT-2 & 0.93 & \underline{\textbf{0.905}} & \underline{\textbf{0.94}} & \underline{\textbf{0.94}} \\\midrule
\multirow{2}{*}{\begin{tabular}{@{}c@{}}(10) Multiple\\discussion channels\end{tabular}}& MDC-1 & 0.53 & 0.545 & 0.545 & 0.545 \\
& MDC-2 & 0.815 & 0.84 & 0.815 & 0.815 \\\midrule
\multirow{3}{*}{\begin{tabular}{@{}c@{}}(11) Insufficient\\documentation\end{tabular}} & ID-1 & 0.765 & 0.775 & 0.755 & 0.75 \\
& ID-2 & 0.81 & 0.795 & 0.79 & 0.79 \\
& ID-3 & \underline{\textbf{0.985}} & \underline{\textbf{0.9}} & \underline{\textbf{0.9}} & \underline{\textbf{0.9}} \\\midrule
\begin{tabular}{@{}c@{}}(24) No task\\progress report\end{tabular} & NTPR-1 & 0.82 & 0.75 & 0.735 & 0.735 \\\midrule
\multirow{2}{*}{\begin{tabular}{@{}c@{}}(28) Unclear\\requirements\end{tabular}} & UR-1 & 0.665 & 0.635 & 0.65 & 0.65 \\
& UR-2 & 0.78 & 0.775 & 0.79 & 0.79 \\\midrule
\multirow{2}{*}{\begin{tabular}{@{}c@{}}(31) Poor onboarding\\process\end{tabular}} & POP-1 & 0.815 & 0.81 & 0.83 & 0.825 \\
& POP-2 & 0.53 & 0.5 & 0.48 & 0.48 \\\midrule
\multirow{2}{*}{\begin{tabular}{@{}c@{}}(42) Inefficient\\meeting system\end{tabular}}& IMS-1 & 0.49 & 0.465 & 0.47 & 0.47 \\
& IMS-2 & 0.665 & 0.675 & 0.67 & 0.675 \\\midrule
\textbf{Average votes} & & 78 & 291 & 413 & 416 \\\midrule
\end{tabular}
\caption{\label{tab:results_tools_ideas_eval}Evaluation of tool ideas across different influence groups: (A)uthority, (I)nfluencer, (S)uggester, (O)bserver.
"Average votes" represents the mean count of total votes across the ten tools for each group combination.
\added{We use micro averaging, giving equal weight to each survey participant.}
We highlight three highest scores for each group.}
\end{table}

%In the survey, we asked each respondent about the influence they have in their teams in regards to adopting new tools.
%We wanted to see if evaluation results would be different for these groups, since opinions from individuals with greater influence are more valuable from an industrial perspective.
%The codes for those influence groups are following:\\
%\textit{Authority}: I can decide that our team will use this tool from now on, and it is unlikely to be overruled by someone else\\
%\textit{Influencer}: I cannot decide that our team will adopt this tool, but I can strongly influence a discussion about whether or not this tool should be used (e.g. I’m one of 3 people who will make this decision together)\\
%\textit{Suggester}: I can suggest using this tool, but I have no impact on whether it will be adopted or not\\
%\textit{Observer}: I have no say in these matters\\

In \Cref{tab:results_tools_ideas_eval}, we present the results of the evaluation of tool ideas across various groups categorized by influence levels. 
For each idea and each group we provide the overall idea score which was calculated with the following formula:
\begin{equation*}
\begin{aligned}
\begin{multlined}
\text{Idea score} = 2 \cdot \text{[ratio of votes for ``Solves the problem completely"]}\;  + \\
    +\;  \text{[ratio of votes ``Solves a part of the problem"]}\;  + \added{0.5 \cdot \text{[ratio of votes ``Helps to manage the problem a little"]}} 
\end{multlined}
\end{aligned}
\end{equation*}
For example, imagine that an idea received 50\% of votes for the  "completely solved" answer, 15\% of votes for "solves a part", 24\% of votes for "manages a little", and 11\% of votes for "not beneficial".
In this case, its idea score will be $2\cdot 0.5 + 0.15 + 0.5\cdot 0.24 = 1.27$.
This scoring allows us to include all respondents who find the ideas valuable. 
Moreover, it emphasises those who perceive a specific idea as capable of resolving the pattern entirely, rather than just partially.
For a comprehensive view of the detailed voting statistics -- percentages of participants who voted for each evaluation answer option within each group -- please refer to Appendix A\footnote{Online Appendix: \url{https://github.com/KatyaKos/Undesirable-Patterns/blob/main/Undesirable_Patterns_Appendix.pdf}}.

\added{It is interesting to note that in most cases, the difference in scores between groups is under 0.7.}
This implies that individuals with various levels of influence generally share a similar perspective on the potential usefulness of the proposed tools.
The scores among the two largest groups (AIS and AISO) remain identical, which is expected since only ten observers (O) participated in the survey.
Nonetheless, certain tool ideas display notable score deviations across groups.
\added{The ideas for the \textit{Uncoordinated code changes} pattern had the highest difference in scores between the authority group and everyone else (0.1-0.14 difference). 
This is intriguing, considering it is a Technical code-related \up; the category of patterns we expect to be easier to identify and address, as it is purely technical.}

Notably, for the majority of tool ideas, the authority group tends to yield the highest score. 
In instances where the authority group's score is lower, the difference is considerably small.
\added{Interestingly, both ideas for the \textit{Outdated tests} pattern receive the lowest idea scores from the authority group.} 
The highest variance in scores is 0.06, it is observed between groups A and AI when evaluating idea \textit{OT-1}. 
This suggests that individuals with the highest influence in introducing new tools to their teams believe in the usefulness of the suggested tools and features, and, consequently, might adopt them in their teams.

\observation{
Out of the 22 tool ideas we evaluate, \added{14 of them receive full idea scores (AISO) equal to or greater than 0.75.
A score higher than 0.75 suggests that the tool has substantial positive feedback.}
In retrospect, results substantiate our selection of patterns and associated tool ideas for evaluation.
Interestingly, ideas for addressing the \textit{Outdated tests} pattern in total received the highest scores, so they might be the most practical to develop.
}

\subsection{Problem Related Statistics}
\noindent
\begin{table}[!htpb]
\small
\begin{tabular}{ccccccc}
\toprule
\textbf{Pattern} & \multicolumn{5}{c}{\textbf{Frequency (\%)}} & \textbf{Total} \\\cmidrule(lr){2-6}
\textbf{name} & \textbf{Often} & \textbf{Sometimes} & \textbf{Exceptionally} & \textbf{Never} & \textbf{N/A} & \textbf{votes} \\\midrule
\rowcolor{gray} \begin{tabular}{@{}c@{}}(4) Negligent\\code reviews\end{tabular} & 27 & 40 & 22 & 8 & 3 & 445 \\
\begin{tabular}{@{}c@{}}(5) Uncoordinated\\code changes\end{tabular} & 13 & 27 & \underline{\textbf{35}} & \underline{\textbf{21}} & 4 & 354 \\
\rowcolor{gray} \begin{tabular}{@{}c@{}}(6) Slow code\\review process\end{tabular} & 30 & 36 & 20 & 11 & 3 & 418 \\
\begin{tabular}{@{}c@{}}(8) Outdated\\tests\end{tabular} & 36 & 31 & 18 & 7 & 8 & 406 \\
\rowcolor{gray} \begin{tabular}{@{}c@{}}(10) Multiple\\discussion\\channels\end{tabular} & 39 & 26 & 18 & 13 & 4 & 393 \\
\begin{tabular}{@{}c@{}}(11) Insufficient\\documentation\end{tabular} & \underline{\textbf{61}} & 29 & 6 & 2 & 0 & 476 \\
\rowcolor{gray} \begin{tabular}{@{}c@{}}(24) No task\\progress report\end{tabular} & 24 & 36 & 26 & 12 & 2 & 433 \\
\begin{tabular}{@{}c@{}}(28) Unclear\\requirements\end{tabular} & 40 & \underline{\textbf{42}} & 12 & 3 & 3 & 422 \\
\rowcolor{gray} \begin{tabular}{@{}c@{}}(31) Poor\\onboarding\\process\end{tabular} & 42 & 25 & 17 & 11 & 5 & 407 \\
\begin{tabular}{@{}c@{}}(42) Inefficient\\meetings system\end{tabular} & 45 & 29 & 11 & 12 & 3 & 399 \\\bottomrule
\end{tabular}
\caption{\label{tab:results_tools_freq}Frequency of 10 patterns reported in the \surveyTwo survey.}
\end{table}

In the \surveyTwo survey, we ask participants about their familiarity with each presented \up.
This step is taken to eliminate participants without relevant experience with the related patterns from the tools evaluation process. 
Given the considerable increase in the number of respondents in the \surveyTwo survey compared to the \surveyOne survey (968 and 436 respectively), we opt to reevaluate patterns' frequency-related statistics, as initially shown in \Cref{sec:results_problems}.

We calculate the average occurrence rate of an \up by dividing the number of respondents who indicate experiencing the pattern (excluding "Never" and "N/A" responses) by the total number of respondents who are suggested to vote on the pattern.
In the \surveyOne survey, the mean average occurrence rate across all \nprobs patterns is 97\%, with the lowest average occurrence rate being 91\%. 
However, if we focus solely on the ten patterns selected for further study, the mean average occurrence rate increases to 98\%, with a minimum of 97\%.
This increase in average occurrence rates explains the higher percentage of respondents who reported encountering all five suggested \ups in the \surveyTwo survey \Cref{tab:results_tools_prob_occurence}. 
This trend is consistent across various participant groups, including management, individuals with less than four years of experience, those with at least seven years of experience, and individuals with significant influence in adopting new tools.

The frequencies of ten patterns reported in \surveyTwo differ from the frequencies in \surveyOne \Cref{tab:results_problems_stats_freq}. 
To check, whether this difference is statistically significant, we do a Mann-Whitney U test~\cite{mann1947test} and use the Holm method~\cite{holm1979simple} to account for multiple comparison.
We use Mann-Whitney U test as the data on frequency does not satisfy requirements for the t-test. 
As we do family-wise comparison, we need to amend p-values corresponding to the statistically significant events to avoid false positives. 
We choose Holm method as a method with higher statistical power than e.g. Bonferroni correction.
We find that for none of the ten patterns we consider the difference in reported frequencies is statistically significant (with family-wise error rate $0.05$).

\noindent
\begin{table}[!htpb]
\small
\centering
\begin{tabular}{cccccccc}
\toprule
 & \multicolumn{6}{c}{\thead{\textbf{Number of problems met}\\\textbf{(out of 5, \%)}}} & \thead{\textbf{Total}\\\textbf{participants}}\\\cmidrule(lr){2-7}
 & \textbf{0} & \textbf{1} & \textbf{2} & \textbf{3} & \textbf{4} & \textbf{5} &  \\\midrule
\rowcolor{gray} \textbf{All} & 1 & 2 & 3 & 12 & 22 & 60 & \nsurveyTwo \\
\textbf{Management} & 2 & 1 & 3 & 8 & 21 & \underline{\textbf{65}} & 307 \\
\rowcolor{gray} \textbf{$<$4 years of experience} & 1 & 4 & 7 & 21 & 32 & 36 & 132 \\
\textbf{7+ years of experience} & 1 & 1 & 2 & 10 & 20 & \underline{\textbf{65}} & 653 \\
\rowcolor{gray} \textbf{High influence} & 1 & 1 & 4 & 12 & 22 & 60 & 677 \\\bottomrule
\end{tabular}
\caption{\label{tab:results_tools_prob_occurence}Percentages of people who met 0-5 \ups out of 5 presented to them in the \surveyTwo survey.}
\end{table}

\FloatBarrier

\section{Discussion and Future Work}\label{sec:discussion}
In this section, we discuss the results of this study, and its implications for research and industry.
We also outline several hypotheses which we derive from our results but are not able to test within this study.
We group the parts of this section according to the key outcomes of our study and future research directions.

\subsection{Undesirable patterns and their properties}
In our \surveyOne survey we gather data on the frequency and impact of \ups, as well as frequency of various consequences.
This data can serve as a rough proxy of the potential monetary costs associated with different \ups, responding to the call to estimate the costs of various community smells by Caballero-Espinoza et al.~\cite{caballero2022community}. 
\added{One could argue that the more significant or impactful \ups require more resources to address.
Regarding frequency, addressing each new instance of an undesirable pattern also necessitates some investment.
However, it is important to note that this cannot replace a comprehensive analysis of monetary costs, and we only suggest that this data can serve as a rough proxy rather than a precise estimation.}
However, there are several directions for the future research that could provide deeper insights into the costs and impact of various \ups.
\\~\\
\noindent
\textbf{1.} We categorize \ups into three distinct groups based on their origins: technical, social, and organizational.
Each of these groups encompasses several types of patterns, as outlined in the typology presented in ~\Cref{sec:results_interview_typology}.
\begin{itemize}
    \item{\textbf{Technical}}: 14 patterns.
    Eight patterns have at least one consequence with occurrence rate above 75\%. 
    Five patterns match with five community smells, two patterns match with two code review smells, and two patterns match with two code smells.
    \item{\textbf{Social}}: 12 patterns.
    Six patterns have at least one consequence with occurrence rate above 75\%. 
    Eight patterns match with ten community smells.
    \item{\textbf{Organizational}}: 17 patterns.
    Eleven patterns have at least one consequence with occurrence rate above 75\%. 
    Six patterns match with ten community smells.
\end{itemize}
\added{All this data can be derived from the Appendixes A.4-6} \footnote{Online Appendix: \url{https://github.com/KatyaKos/Undesirable-Patterns/blob/main/Undesirable_Patterns_Appendix.pdf}}.

\textbf{Technical patterns} do not exhibit dominance in any regard: they account for roughly one third of the \ups we identify.
Their frequency, significance and number of consequences do not surpass those of other types. 
Nevertheless, technical \ups appear to have garnered more attention in research. 
For example, the share of code, code review or community smells matching with various technical \ups is the highest of the three groups we consider. 
This preference for technical patterns can be due to the "availability heuristic".
Technical \ups are often more straightforward to describe, and they are easier to address with tools.
\added{A lot of them have been addressed long ago and have been discussed in literature for years}~\cite{techi1, techi2, techi3}.
Our study shows that out of the ten \ups accompanied by viable tool suggestions that we investigate in \surveyTwo, six belong to the technical group.

\textbf{Social patterns} appear to be the most notable in the community smells (eight patterns to ten smells).
\added{There are works that explore social aspects of software development and how they influence developers, productivity, and other factors}~\cite{social1, social2, social3, social4}.
However, there appear to be relatively few tooling solutions available for this particular group.
For example, only one of the ten matching community smells, the \textit{Lone Wolf} smell, has been extensively studied~\cite{caballero2022community}.
Moreover, tool suggestions for addressing social \ups are comparatively scarcer than those for other groups.
We could select only one tool targeting a social pattern to feature in the \surveyTwo. 
We believe that an exhaustive study on strategies to address social \ups would help software engineering practitioners. 

\textbf{Organizational patterns} appear to be the most important, according to the \surveyOne survey.
As we mention in \Cref{sec:results_problems_stats}, most patterns in the top ten most frequent and top ten most significant lists are organizational patterns.
A portion of these organizational \ups generalizes and encompasses several recognized community smells (six to ten).
\added{There is a number of researchers who study the way organizational structures and policies impact software development}~\cite{organiz1, organiz2, organiz3, organiz4}.
Unfortunately, the adoption of solutions proposed by the research community has been challenging for practitioners.
However, an extensive investigation into viable strategies to mitigate organizational \ups extends beyond the scope of this study while being an interesting direction for future work.
\\~\\
\noindent
\textbf{2.} In our surveys, we choose not to inquire why the specific \up encountered by respondents was left unaddressed.
Posing an open-ended question can potentially yield a low number of responses. 
Obtaining such data through interviews would require conducting numerous interviews for each pattern, resulting in an unfeasible workload.
Investigating the reasons behind the lack of resolution for each \up can help quantify the impact of various decisions leading to these \ups. 

In our interviews, we ask respondents about the specific reasons behind cases where an \up was left unattended.
Interestingly, many respondents mention the lack of resources as a primary factor.
This finding agrees with a previous study~\cite{jabrayilzade2022bus} and highlights the importance of putting a "price tag" on these \ups for software engineering practitioners.
In many cases, the allocation of resources to a team is determined by upper management. 
Thus, assigning a measurable value to an \up could enable team manager to perform a cost-benefit analysis.
This analysis could convince higher management to allocate more resources to address these challenges effectively.
\\~\\
\noindent
\textbf{3.} We find that the IT practitioners with seven or more years of experience have witnessed more \ups, than the ones with less than four years of experience.
This supports the claim made by Caballero-Espinoza~\cite{caballero2022community} that individuals with more experience are more likely to notice \ups. 
However, we did not study what enables practitioners to recognize an \up, and recognizing a problem is necessary to address it.
Therefore, we consider it important to investigate the characteristics of practitioners that influence their ability to identify \ups.
Such an investigation could hold significant value for both the research community and practitioners.
\\~\\
\noindent
\textbf{4.} While our study examines the frequency with which each \up results in a particular consequence, we do not explore the reverse relationship.
Analyzing how often each undesirable pattern becomes the primary cause for a given consequence would be valuable. 
This is especially important because different consequences can vary significantly in terms of their relative costs for different software engineering projects. 
Understanding the relationship between consequences and patterns could provide IT practitioners with insights on which patterns to prioritize in order to mitigate certain risks. 
Unfortunately, we cannot easily obtain this data from our study results, as more than one \up can lead to a particular consequence.
\\~\\
\noindent
\textbf{5.} We hypothesize that the co-occurrence of different \ups depends on their origins, with \ups of similar origins being more likely to co-occur.
To illustrate this, let's consider two pairs of \ups.
One pair consisting of two technical code-related \ups, namely \textit{Breaking change} and \textit{Outdated tests}, and another pair containing \textit{Outdated tests} and \textit{Ignoring directions} from the social reluctance category.
We speculate that the patterns within the first pair tend to co-occur more frequently compared to those in the second pair. 
Unfortunately, the sample size of respondents in the \surveyOne survey does not allow us to draw statistically significant conclusions on this matter.
However, if our hypothesis holds true, it could carry significant research and practical implications.

From the research perspective, if a group of \ups frequently co-occur, it could suggest an underlying common issue that might not be evident to practitioners. 
Investigating why these patterns tend to co-occur could improve our understanding of how software engineering teams collaborate.
From the practical perspective, understanding whether a hard-to-identify pernicious \up often co-occurs with a more recognizable \up would be valuable. 
This could enable the practitioners to assess whether their project might have a challenging-to-detect yet impactful \up.
\\~\\
\noindent
\textbf{6.}
\added{In this research, we do not study the relationship between the frequency or impact of \ups and the general properties of the company and the team in which the respondents work (such as company size, or development type). 
We choose not to inquire about the properties of the company where our survey respondents work. 
This decision stems from the fact that the number of responses for each pattern in the \surveyOne survey was approximately 50, see}~\Cref{tab:results_problems_stats_freq}. 
\added{Dividing this sample into multiple groups would yield statistically insignificant results for at least some of the groups. 
Furthermore, as the surveys are already extensive, adding more questions would likely result in a decline in responses quality}~\cite{survey1, survey2}
\added{Nonetheless, we believe that it is worthwhile to study whether the properties of the team and the company affect the \ups the team or the company faces.}

\added{For the same reasons, we do not inquire about participants' personal characteristics.
It is established that developer's personality affects their problem-solving skills, focus, adaptability, and overall performance}~\cite{personality1, personality2, personality3}.
\added{It would be interesting to see if there any correlations between \ups and personality traits.}
\\~\\
\noindent
\textbf{7.} Finally, it is interesting that several community smells, such as \textit{Unlearning} and \textit{Instutional Isomorphism}~\cite{caballero2022community}, do not have any corresponding \ups. 
This could mean that the practitioners fail to recognize these community smells and observe their distant consequences instead.

\subsection{Tools for undesirable patterns}
In this study we focus on addressing \ups with tools, as using a tool is cheaper and easier than changing the policy in a team or organization.
However, there are several possible issues with addressing \ups with a tool.
\\~\\
\noindent
\textbf{1.} Not all \ups can be effectively addressed with tools. 
Out of \nprobs \ups we collect, ten lacked any tool suggestions from our interview respondents. 
While creating tools might be feasible for some of these \ups, others, such as \textit{Toxic company culture}, cannot be reliably found or managed with a tool.
Even defining the presence of such \ups is highly subjective and dependant on individual employee characteristics.
In other cases tools can provide partial solution, yet effective resolution often demands additional team actions. 
For example, when dealing with \textit{Informational bottleneck}, a tool might identify the problem and recommend adding more people to the bottlenecked part of the project.
However, it cannot autonomously spread project knowledge.
Therefore, we believe that creating organizational strategies or guidelines to address \ups can be beneficial for practitioners.
However, this research should consider cost-benefit analyses. 
A convenient albeit imperfect guideline is likely more useful in practice, than a exhaustive yet cumbersome one.
\\~\\
\noindent
\textbf{2.} Several ideas proposed by our respondents already exist as research prototypes or even market tools, yet software companies have been hesitant to adopt them.
For instance, the tool proposal \textit{POP-1} closely resembles StackOverflow for Teams \cite{stackoverflowTeams}, a product that has been on the market since 2018.
We hypothesize that the lack of adoption can be attributed to three potential factors:
\begin{itemize}
    \item \textbf{People don't know about the product.} 
    This challenge is most prominent for tools originating from the research community. 
    Nonetheless, circumstantial evidence, as seen with StackOverflow, suggests that even well-established companies might not possess full brand awareness.
    \item \textbf{Required costs outweighing benefits.}
    Adopting new functionality requires certain resources, even if it doesn't involve direct monetary costs.
    The company must handle the maintenance of each tool, tackle potential compatibility problems (which in themselves can be undesirable patterns), and manage associated security risks. 
    Teams need to incorporate the tool into their existing workflow and provide training to employees for proper tool usage.
    These costs might exceed the anticipated benefits derived from employing the new tool.
    %\item \textbf{Lack of resources.}
    %Adopting new functionality requires certain resources, even if it doesn't involve direct monetary costs.
    %Teams need to incorporate the new feature into their existing workflow and provide training to employees for proper tool usage.
    %This process can be challenging, particularly when team members are time-constrained.
    %\item \textbf{Unwillingness to add more tools to the stack.}
    %Incorporating a new tool into the stack introduces additional costs compared to utilizing a feature of an existing tool. 
    %The company must handle the maintenance of each tool, tackle potential compatibility problems (which in themselves can be undesirable patterns), and manage associated security risks. 
    %These costs might exceed the anticipated benefits derived from employing the new tool.
    \item \textbf{Dislike of the tool.}
    Certain patterns could prove difficult to identify before they escalate into serious issues. 
    Sometimes it might be challenging to understand or believe that a situation could lead to undesirable consequences.
    If a tool lacks appeal or convenience for users, employees might be reluctant to use it if they do not perceive potential risks as substantial or significant.
\end{itemize}
All these hypotheses can be briefly summarized as "the tool with the most significant impact on an \up is a user-friendly plugin or feature that can be easily adopted within a widely-used software platform".
Indeed, many research prototypes fall short in meeting one or more of the criteria we have outlined.
Ideally, any tool prototype should be accompanied by a study that assists its creators in identifying the target audience and designing a user-friendly tool.
\\~\\
\noindent
\textbf{3.} In the \surveyTwo survey, we ask respondents about their influence on tool adoption decisions.
We then examine whether opinions on particular tool ideas vary based on the respondent's level of authority.
However, only ten respondents identify themselves as Observers, or individuals with no influence whatsoever.
Given the small number of Observers, it might be beneficial to conduct an additional study to gather opinions from various influence groups regarding existing and potential tools.
In our study, in the \surveyTwo survey, the Authority group is generally the most enthusiastic to adopt a tool.
Still, for the majority of ideas, the difference in opinions between decision-making respondents and those without such authority is not substantial.
We believe it is necessary to study whether Authority tends to compel other employees to adopt specific tools or practices.
From a practitioner's perspective, it is advisable for a manager to ensure with the team that the decisions made align with their needs and preferences.
\\~\\
\noindent
\textbf{4.}
\added{The interest in the tools we describe in the \surveyTwo survey might not necessarily translate to actual market interest if these tools are to be developed.
One important reason for this possible discrepancy is the gap between the expressed interest of respondents who were intrigued by a tool's description and the real-world enthusiasm for an actual prototype.
Overcoming this potential issue would require conducting user studies for each proposed tool, which is beyond the scope of this study.}
\section{Threats to validity}\label{sec:validity}
\subsection{Construct validity}\label{sec:validity_construct}
Construct validity threats arise from the relationship between theory and observations.
In our study, the primary source of construct validity threats originates from interpreting interview data and processing it for the surveys.
\\~\\
\noindent
\textbf{Threat 1.}
\added{The first threat to construct validity arises from the use of surveys and interviews to discover and analyze \ups.
While surveys and interviews can help with understanding people's attitudes (what people say), the \ups that occur in reality stem from people's behaviour (what people do). 
The correlation between attitudes and behaviour has been a longstanding issue in sociology~\cite{weinstein1972predicting}.
Unfortunately, conducting a behavioural study with statistical power and external validity comparable to ours would require longitudinal observation of multiple teams across various companies. 
However, such a study is not feasible within our budget constraints and would likely encounter self-selection bias. 
Longitudinal team observation requires close researcher involvement, which may not be feasible for all teams due to discomfort or legal concerns such as intellectual property protection.}

\added{As direct study of IT practitioners' behaviour is unfeasible, to mitigate the threat to construct validity, we need to formulate the surveys and interviews in a way that the studied attitudes will have a strong correlation with actual behaviours. 
For example, when studying the impact of an \up, we need to focus the respondent on the work-related impact, rather than a general "how bad it is" sentiment.
To achieve this, we follow the meta-analysis of Glasman and Albarracin on attitude-behavior relation~\cite{glasman2006forming} and ground our respondents in the recent experiences of the \ups they have faced. 
By discussing recent experiences, interviews can easily recall specific situations and their actions, making their attitudes more strongly connected with actual behaviours. 
This approach is applied consistently to all \up properties we analyze.}

\added{For RQ3, we gathered tool ideas from interviewees and then asked survey respondents to evaluate these suggestions.
This research question focuses on attitudes that cannot be grounded in practical experience (as it is impossible to evaluate a non-existing tool in practice).
Therefore, the attitude-behaviour link is likely to be weaker, and we only present the respondents' opinions on the tool ideas.
Acknowledging these limitations, the previous section outlines possible reasons why positive feedback for the tool ideas may not translate into interest in the real-world prototypes.}
\\~\\
\noindent
\textbf{Threat 2.}
Different interpretations of \ups could lead to some of the identified patterns being irrelevant.
Moreover, the descriptions of these patterns presented in the survey might be misleading, causing respondents to misinterpret them and not understand the original intent.

To address this concern, the first two authors of this study take several steps. 
We independently classify the \ups and engaged in multiple discussions to reach a consensus on the final list of patterns.
Then, we agree on the definitions of the patterns and use the help of two experienced proofreaders for further validation. 
Finally, we test each definition through interview-like discussions with several IT practitioners to ensure clarity and accuracy.
During these sessions, a practitioner would read the description of a pattern and explain the content in their own words.
Then, we verify if the explanation matches the original intent. 
This process is repeated for each definition, question, and description, involving at least two IT practitioners for each.
Interestingly, during this testing, certain \ups that might seem closely related, like \textit{Negligent code review} and \textit{Slow code review process}, were identified as distinct by the interviewees.
This distinction is often rooted in the respondents' practical experience.

To further address potential validity threats stemming from this source, we include a comment section in the \surveyOne survey.
This section follows each \up part of the survey and encourages respondents to provide their opinions on the patterns they evaluated.
Notably, we receive no complaints or concerns about the definitions and descriptions of the patterns in this comment section.
\\~\\
\noindent
\textbf{Threat 3.}
The classification of types and consequences of the \ups we present in this study also introduces a potential threat to construct validity.
To mitigate this threat, we adopt procedures similar to those explained in Threat 1.
Both authors of this study independently categorize the types and consequences of the undesirable patterns.

Moreover, in the \surveyOne survey, questions about the consequences of \ups include an open option that allows respondents to add any other consequences they may have observed. 
We adopt a similar approach for defining tools, incorporating an open-ended option to gather any additional insights or suggestions from the respondents.
This strategy enhances the comprehensiveness and accuracy of the data collected in the study.

\subsection{Internal validity}\label{sec:validity_internal}
Certain survey participants might lack relevant experience to answer questions adequately.
To address this, we exclude responses from those who had never worked in the IT field.

Another concern is that \ups correspond to teamwork problems.
Respondents might be hesitant in sharing their opinions due to the perceived intrusive nature of the questions.
While we attempt to mitigate this by assuring that all survey and interview results would be anonymized and presented in an aggregated manner, complete resolution may not be achievable.

\subsection{External validity}\label{sec:validity_external}
%External validity threats are related to the generalization of the results we obtained.
The interviews we conduct for this study involve participants from 17 different countries, ensuring a diverse range of perspectives.
Each job role, level of experience, and type of software development (such as startups and in-house tools) are represented by at least 5 respondents. 
To enhance the generalizability of our findings, we recruit interview and survey participants through mailing lists associated with \jb, which has a substantial number of entries.
We collected 436 responses for the \surveyOne survey and 968 responses for the \surveyTwo survey.
While we exhaust our options in recruiting survey participants, we acknowledge that the list of \ups we compiled might not be comprehensive.

\added{The best solution would be to attract more participants on each study phase.
However, we sent more than ten thousands invitation emails on each phase, and we followed the best practices, including reward for completion, authority, time limit and flexibility}~\cite{survey1}.
\added{The other approach would be to include more questions into our interviews and surveys, which would allow us to (1) present more \ups to each participant, (2) collect more information, e.g. development type or details on patterns.
However, our interviews and surveys were already rather extensive (90 minute interviews, 20 minute surveys), and the more questions are asked the more time it takes to complete it.
If the respondents consider this time as too high they will not finish answering, and the quality of replies often reduces with time}~\cite{survey2}.
Future experiments and replications of our study could potentially uncover more such patterns, thus contributing further to the community's knowledge.

One of the important limitations of this study is that we conduct all our surveys and interviews in English.
Therefore, our study leaves out IT practitioners who are not fluent in English or do not want to use it.
It would be interesting to see extensions to this study with surveys and interviews conducted in other languages.

A recurring study of this nature in subsequent years might yield variations in the collected \ups.
Our interviews were conducted in 2022, with respondents reflecting on experiences from the previous year.
This focus on events from 2021 or 2022 is sensible given the potential for individuals to forget important details of the past events.
However, it is important to recognize that our study's timeline coincides with the global impacts of the COVID-19 pandemic.

\subsection{Ethical Considerations for Tool Usage and Adoption}\label{sec:validity_ethics}
Some of the patterns identified in this study, such as \textit{No Task Progress Report}, could be loosely categorized as issues related to productivity.
Users might consider tools addressing these patterns as proxies for evaluating an engineer's productivity.
However, we assert that this is not an appropriate way to utilize any automated tool, including the ones we propose in our study.

The first concern lies in the likely backlash that can arise from using a tool to assess an engineer's performance and subsequently making decisions based on the tool's output. 
An illustrative example is the XSolla case~\cite{xsolla}, where a group of employees was laid off based on "big data analysis of their productivity".
This case garnered significant negative attention for the company, decreasing trust within the engineering community.

Secondly, engineers in general may be wary of new tools within a company.
For instance, Jabrayilzade et al.~\cite{jabrayilzade2022bus} reported respondents expressing concerns during a survey on the bus factor tool they conducted.
The engineers were concerned about concepts like "bus factor" and "key engineer," even though the tool proposed in that study was not intended as a measure of individual performance.
Introducing a tool explicitly designed to evaluate an engineer's performance could potentially undermine employees' trust in other tools that are not meant for that.

Finally, we believe that any tool, regardless of its purpose, is error-prone.
Therefore, the results produced by a tool should be regarded as just one of the various information sources used to make decisions, rather than being considered a definitive metric of a project or team.
\section{Conclusions}\label{sec:conclusions}
Prior studies addressing teamwork issues tend to approach these problems through specific lenses (e.g., socio-technical frameworks).
While such approaches allow to build models that explain the emergence and impact of specific problems within teamwork, they might inadvertently overlook certain issues.
\added{For example, the literature we are aware of does not mention any problems that would be similar to such \ups as Breaking change or Teamwork pipeline bottleneck.}
This study is the first extensive study of \ups in teamwork within the software engineering domain.
It catalogues a diverse range of problems that can potentially arise in a software engineering team.
We identified \nprobs distinct \ups and established connections between them and existing community, code, or code review smells.
Using interview and survey results, we formulated a system of consequences that can arise from each \up.
Additionally, we established a broad typology that outlines the various origins of these patterns.

In the \surveyOne survey, we engaged practitioners to assess the impact and frequency of the \ups we uncovered during the interviews.
This yielded data that characterizes the varying impact of these patterns.
Using the survey results along with the typology of \ups origins, we determined that the research community tends to have less awareness about social and organizational patterns and potential strategies for addressing them.
Additionally, we validated the classification of consequences and collected data concerning the correlation between patterns and their associated consequences.

Moreover, during the interviews, we gathered an extensive collection of tool and feature ideas suggested by practitioners as potential full or partial solutions for addressing various \ups.
For the \surveyTwo survey, we selected ten patterns with the most promising tool and feature ideas.
We then analyzed the attractiveness of these proposals to practitioners.

In summary, our paper contributes to the field in several significant ways.

\observation{
\textbf{RQ 1.}
We engaged IT practitioners to compile an extensive list of \ups in collective development that they encounter in their professional practice.
This list provides insight into the pain points of real-world software development teams.
Some of the patterns in our list, up to the best of our knowledge, have not been previously studied. 
This highlights the novelty of our research and the need for additional investigation in these areas.
}

\observation{
\textbf{RQ 2.}
We enriched the list of \ups with information about the impact of each pattern, which holds value for both researchers and practitioners. 
Researchers can utilize this list to focus on studying particularly impactful \ups.
Practitioners can identify and address the specific patterns that are most harmful to their teams or organizations.
}

\observation{
\textbf{RQ 3.}
We conducted an evaluation of tool proposals for ten of the identified \ups, each with what we perceived to be the most viable and innovative tool and feature concepts from an industrial perspective.
This list is not intended as a straightforward feature request, as we elaborate in \Cref{sec:discussion}.
However, we consider this study as a foundational step in the exploration of these proposed tools.
This evaluation can serve as a preliminary filter for selecting which tools should be further developed.
The practitioners' insights provide valuable guidance in understanding the practical relevance and potential impact of these tools, enabling a more focused and informed approach in the development process.
}

Since we did not intend for this study to comprehensively cover all \ups and their attributes, we anticipate that future research will address the ideas and hypotheses we could not fully investigate in this study.
\section*{Acknowledgements}\label{sec:acknowledgement}
We are grateful to the \jb Market Research \& Analytics team for providing resources and assistance that facilitated the smooth progress of this research.
Special thanks to Olga Vorobeva, Elli Ponomareva, and Anna Kadnikova who helped us with participant recruitment process, assisted with the first interviews, and reviewed our interview and surveys scripts.
Their input was instrumental in shaping the research.

We would also like to extend our gratitude to Mikhail Melnikov and Andrew Kozlov who devoted their time and expertise to the trial runs of the surveys. 
Their valuable feedback and insights greatly contributed to the refinement and effectiveness of the survey phases of our study.

We want to extend our appreciation to Pouria Derakhshanfar, Sergey Titov, Agnia Sergeyuk, and Olga Vorobeva who dedicated their time and expertise to thoroughly review this paper. 
Their constructive feedback and insightful suggestions greatly contributed to enhancing the clarity and quality of the content. 
We are grateful for their meticulous attention to detail and valuable ideas, which have undoubtedly improved the overall presentation of this work.
We are also grateful to Yaroslav Golubev for reimagining the illustrations and enhancing the visual appeal of the tables that has played a significant role in elevating the overall presentation of this paper. .

We would like to express our sincere gratitude to the research participants who generously shared their time and experiences, contributing immensely to the depth and richness of this study. 
Their willingness to engage in this research is deeply appreciated.

We are also grateful to the TOSEM Editorial Board for their time and thoughtful critique of our work.
Their constructive suggestions have greatly contributed to enhancing the quality and comprehensiveness of this paper.

\bibliographystyle{IEEEtran}
\bibliography{patterns}
\end{document}